\documentclass[aps,groupedaddress,superscriptaddress,amsmath,amssymb,pra,twocolumn,longbibliography]{revtex4-2}
\usepackage{dcolumn}
\usepackage{bm}
\usepackage[hidelinks]{hyperref}
\usepackage{float}
\usepackage{graphicx}
\usepackage{amsfonts}
\usepackage{verbatim}
\usepackage{bbm}
\usepackage{braket}
\usepackage{qcircuit}
\usepackage[normalem]{ulem}
\usepackage[T1]{fontenc} 
\hypersetup{
    colorlinks=true,
    linkcolor=blue,
    filecolor=magenta,      
    urlcolor=cyan,
}
\usepackage{color}
\usepackage{multirow}
\usepackage{array}
\usepackage{tabularx}
\renewcommand{\figurename}{Figure}

\def\sgn{\mathrm{sgn}}

\def\ep{\varepsilon}

\newcommand{\sumprime}[1]{\sum_{#1}{\vphantom{\sum}}^{\!\!\prime}}

\def\h1{\mathds{1}}

\draft

\begin{document}
%\linenumbers
\title{Controlled asymmetric Ising model implemented with parametric micromechanical oscillators}
\author{C. Han}
\affiliation{Department of Physics, the Hong Kong University of Science and Technology, Clear Water Bay, Kowloon, Hong Kong, China}
\affiliation{William Mong Institute of Nano Science and Technology, the Hong Kong University of Science and Technology, Clear Water Bay, Kowloon, Hong Kong, China}
\author{M. Wang}
\affiliation{Department of Physics, the Hong Kong University of Science and Technology, Clear Water Bay, Kowloon, Hong Kong, China}
\affiliation{William Mong Institute of Nano Science and Technology, the Hong Kong University of Science and Technology, Clear Water Bay, Kowloon, Hong Kong, China}
\author{B. Zhang}
\affiliation{Department of Physics, the Hong Kong University of Science and Technology, Clear Water Bay, Kowloon, Hong Kong, China}
\affiliation{William Mong Institute of Nano Science and Technology, the Hong Kong University of Science and Technology, Clear Water Bay, Kowloon, Hong Kong, China}
\author{M. I. Dykman}
\email{dykmanm@msu.edu}
\affiliation{Department of Physics and Astronomy, Michigan State University, East Lansing, Michigan 48824, USA}
\author{H. B. Chan}
\email{hochan@ust.hk}
\affiliation{Department of Physics, the Hong Kong University of Science and Technology, Clear Water Bay, Kowloon, Hong Kong, China}
\affiliation{William Mong Institute of Nano Science and Technology, the Hong Kong University of Science and Technology, Clear Water Bay, Kowloon, Hong Kong, China}

\date{\today}

\begin{abstract} 
Asymmetric Ising model, in which coupled spins affect each other differently, plays an important role in diverse fields, from physics to biology to artificial intelligence. We show that coupled parametric oscillators provide a well-controlled and fully characterizable physical system to implement the model.
Such oscillators are bistable. The coupling changes the rate of interstate switching of an oscillator depending on the state of other oscillators. Our experiment on  two coupled micromechanical resonators reveals unusual features of  asymmetric Ising systems, including the onset of a probability current that circulates in the stationary state. We relate the asymmetry to the exponentially strong effect of a  periodic force on the switching rates of an individual parametric oscillator, which we measure. Our findings open the possibilities of constructing and exploring asymmetric Ising systems with controlled parameters and  connectivity.

\end{abstract}

\maketitle

\section{Introduction}
\label{sec:Intro}

Parametric oscillator is one of the best-known examples of a bistable system. It has two vibrational states with equal amplitudes and opposite phases \cite{Landau2004a}. These states emerge when the oscillator eigenfrequency is periodically modulated. They have a period equal to twice the modulation period and can be associated with classical bits or Ising spin states, providing a basis for classical logic operations~\cite{Goto1959,Mahboob2008a}. Superpositions of the opposite-phase coherent states of an oscillator can also encode a qubit \cite{Leghtas2015,Grimm2020}. Coupled parametric oscillators can serve as  Ising machines for classical and quantum annealing \cite{Wang2013,McMahon2016,Goto2016,Puri2017,Goto2018, Bello2019,Yamamoto2020,Heugel2022a,Alvarez2023}. Besides computation, various other applications of parametric oscillators have been studied, from force and mass sensing \cite{Rugar1991,Karabalin2011} to rare events in classical and quantum systems far from thermal equilibrium \cite{Dykman1998,Lapidus1999,Marthaler2006,Chan2007,Chan2008a,Venkatraman2022a} and phase transitions into a time-symmetry-broken (time-crystal) state \cite{Kim2006,Dykman2018,Heugel2019a}. 

An important aspect of Ising systems pointed out by Hopfield \cite{Hopfield1982} is the possibility to use coupled spins to model neural networks which memorize multiple patterns.  This possibility has been attracting increasing interest over the years, particularly in view of the progress in machine learning \cite{Krotov2023,Lucibello2023}. In the Hopfield model the spin coupling energy has the conventional form of $J_{ij} \sigma_i\sigma_j$, where $\sigma_i, \sigma_j$ take on values $\pm 1$ and $J_{ij}=J_{ji}$, and the network dynamics can be analyzed using the methods of statistical physics. The model is symmetric in the sense that the effect of spin $i$ on spin $j$ is the same as the effect of spin $j$  on spin $i$.

However, most neuron networks are presumably asymmetric: neuron $i$ can affect neuron $j$ stronger than neuron $j$ affects neuron $i$. If neurons are associated with spins, one can think formally that $J_{ij}\neq J_{ji}$ and then the coupling may not be described  by the coupling energy. The corresponding model is called an asymmetric Ising model. It has attracted much attention as one of the leading models of neural networks  \cite{Sompolinsky1986,Parisi1986,Derrida1987,Mezard2011,Huang2014,Aguilera2021} and  gene regulatory networks \cite{Szedlak2014} and has been used to describe experiments on neurons, cf. \cite{Cocco2009,Mendoza2022} and references therein.

In spite of the importance of the asymmetric Ising model, there have been no studies that  relate  the spin coupling parameters to  the parameters of the underlying system. Understanding the dynamics of this system enables one to examine to what extent the mapping on coupled spins is adequate, in the first place. Determining  the relationship between the parameters of the system and the effective spins is essential for implementing and exploring asymmetric Ising models. 

In the present paper we demonstrate that coupled parametric oscillators in the presence of noise provide a system that can be described by an asymmetric Ising model. The description is based on the Glauber picture \cite{Glauber1963} in which the rate of switching between the  states of a spin depends on the states of the spins to which it is coupled. In the case of oscillators, the relevant quantity is the rate of switching between the period-two vibrational states of an oscillator that depends on which vibrational states are occupied by other oscillators.  We describe the mapping of the oscillators on spins and independently measure the parameters of the system that enter the model. In particular, we measure an important characteristic of driven oscillators in the presence of fluctuations, the logarithmic susceptibility \cite{Smelyanskiy1997b}, which describes the exponentially strong effect of a periodic force on the switching rates of an individual uncoupled parametric oscillator. 

The parametric oscillators we study are micro-electro-mechanical resonators  modulated close to twice their eigenfrequencies. Such resonators enable exquisite control of their eigenfrequencies and the coupling. Since the decay rates of our resonators are small, the modulation needed to excite parametric vibrations is comparatively weak, so that the vibrations are nearly sinusoidal.

With micromechnical resonators, we demonstrate  that the asymmetric Ising model does not have detailed balance. An immediate consequence is the emergence of a probability current that circulates in the system in the stationary state. We measure this current for a system of two coupled non-identical parametric oscillators. The measurements are in excellent agreement with the theory. 

We consider the case where the coupling of the oscillators is weak, so that each oscillator still has two stable vibrational states, and their amplitudes and phases are only weakly changed by the coupling. However, the coupling can significantly change the rates of noise-induced switching between the states. To gain an intuitive understanding, consider a Brownian particle in a  symmetric double-well potential. Because of thermal fluctuations, the particle switches between the wells with the rate $W\propto \exp(-\Delta U/k_BT)$, where $\Delta U$ is the barrier height and $T$ is temperature \cite{Kramers1940}. If the potential is tilted, the barrier heights are incremented by $\pm \delta U$ in the opposite wells, breaking the symmetry of the interwell switching rates. The rates acquire extra factors $\exp(\pm \delta U/k_BT)$. Even for a small tilt, the ratio $\delta U/k_BT$ can be large, for low temperatures. In that case the stationary populations of the wells  become significantly different. 

Consider now a set of weakly interacting particles in double-well potentials. A particle exerts force on other particles that depends on which well it occupies. This force tilts the potentials of the other particles and breaks the symmetry of the interwell switching rates, reminiscent of the effect of the spin-spin coupling in the Glauber model. The change of the switching rates  of the spins is fully determined by the coupling energy,  which in turn depends only on the relative spin orientations. For example, for two coupled spins, the change $\propto \exp(-J_{12}\sigma_1\sigma_2/k_BT)$ is the same for both of them. In other words, the two spins affect each other symmetrically. As we show, the picture extends to coupled parametric oscillators, even though there are no static double-well potentials. However, a major difference is that the oscillators can affect each other asymmetrically. 

If the oscillators are identical and the coupling is weak,
the changes of the switching rates are equal within each pair of coupled oscillators. The system is mapped onto the symmetric Ising model. On the other hand, if the oscillators have different parameters, we show that the coupling-induced changes of the switching rates are different. The picture of a change in potential barriers no longer applies. Instead, the system is mapped onto the asymmetric Ising model. In our  system, switching between the period-two vibrational states is activated by noise with controlled intensity, which allows us to fully characterize the switching rates.

\section{Results}
\label{sec:results}

We present experimental results for a system of two micromechanical torsional resonators. They are shown in Fig.~1a. Each resonator consists of a movable polysilicon top plate ($200\,\mu$m$\,\times\,200\,\mu$m$\,\times\,3.5\,\mu$m) supported by two torsional rods, with two fixed electrodes underneath. The resonators are located side by side. Their vibrations can be excited and detected independently.  For resonator $i$ ($i = 1, 2$), dc voltages  $V^{\mathrm{dc}}_{\mathrm{L},i}$, $V^{\mathrm{dc}}_{\mathrm{R},i}$ and $V^{\mathrm{top}}_i$ are applied to the left electrode, the right electrode and the top plate respectively. Application of an ac voltage on the left electrode generates a periodic electrostatic torque that excites vibrations of the top plate. The vibrations are detected by measuring the current flowing out of the top plate induced by the capacitance change between the plate and the two underlying electrodes.

\begin{figure}
%for reprint
\includegraphics[scale=0.4]{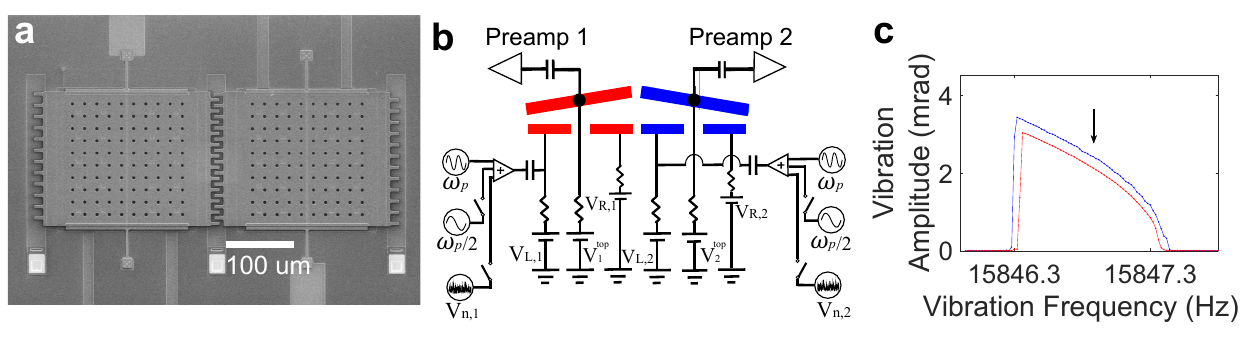}
%for preprint
%\includegraphics[scale=0.8]{Fig1.pdf}
\caption{\textbf{Two coupled parametric torsional resonators. a.} Scanning electron micrograph of two torsional resonators located side-by-side. The scale bar measures 100 $\mu$m. \textbf{b.} Schematic of the actuation scheme and measurement circuitry. Voltages applied to the left electrodes generate the parametric modulation at $\omega_p$, the drive at $\omega_p/2$ and the noise. Voltages $V^{\mathrm{dc}}_{\mathrm{R},i}$ applied to the fixed electrodes allow fine tuning of the resonant frequencies. The dc voltage differences between each top plate and the underlying electrodes leads to an ac current flowing out of the top plate as it rotates. Capacitive coupling between the two plates is controlled by the voltage difference $V^{\mathrm{top}}_1 - V^{\mathrm{top}}_2$. \textbf{c.} Vibration amplitude of resonators 1 (red) and 2 (blue) subjected to identical parametric modulation as functions of $\omega_p/2$. There is no coupling between the resonators and the eigenfrequencies are tuned to be almost equal. 
}
\label{fig:1} 
\end{figure}

In this study, only the fundamental modes of torsional vibrations are used. The eigenfrequencies of the resonators are almost identical, with $\omega_1/2\pi  \approx 15860.562$~Hz for resonator 1 and $\omega_2/2\pi\approx 15860.598$~Hz for resonator 2; they  can be fine tuned by adjusting dc potential difference $\Delta V_{\mathrm{R},i} = V^\mathrm{top}_i - V^{\mathrm{dc}}_{\mathrm{R},i}$ between the plate and the corresponding  right electrode (Supplementary Note 1). The damping constants are $\Gamma_1/2\pi\approx  0.064$~Hz and $\Gamma_2/2\pi \approx 0.063$~Hz for resonators 1 and 2 respectively.  

The spring constants of the both resonators are modulated electrostatically together at frequency near $2\omega_{1}\approx 2\omega_2$, leading to parametric excitation of the vibrations. We also inject broadband Gaussian voltage noise for each resonator that leads to occasional switching between the period-two vibrational states. 

As shown in Fig.~1a, the adjacent edges of the plates form interdigitated comb-shaped electrodes to allow the plates to couple electrostatically when there is a potential difference $V_\mathrm{cpl}= V^{\mathrm{top}}_1 - V^{\mathrm{top}}_2$ between them. When $V_\mathrm{cpl}$ = 0 V, we verify that there is no coupling between the plates. We keep  $V_\mathrm{cpl}$ small as we focus on the regime of the weak coupling that only weakly perturbs the dynamics in the absence of noise.

All measurements are performed at room temperature at pressure below 10 $\mu$torr. The eigenfrequencies and the coupling  between the resonators can be tuned independently (Supplementary Note 1), which is crucial for revealing the features of the asymmetric Ising model. 

The equations of motion of coupled parametric oscillators have the form

\begin{align}
\label{eq:eom}
&\ddot q_i + 2\Gamma_i \dot q_i + \omega_i^2 q_i  + \gamma_i q_i^3 + M_i^{-1}\sumprime{j}V_{ij}q_j\nonumber\\
&=(F_p/M_i)q_i\cos\omega_pt + \xi_i(t) .
\end{align}
For our pair of torsional resonators, $i=1,2$. The coordinate $q_i$ is the rotation angle of the $i$th resonator,  $M_i$ is its moment of inertia, $\gamma_i$ is the Duffing nonlinearity parameter, $F_p$ and $\omega_p$ are the amplitude and frequency of the parametric modulation, respectively, and $\xi_i(t)$ is zero-mean Gaussian noise of controlled intensity $4D_i\Gamma_i$, $\braket{\xi_i(t)\xi_j(t')} = 4D_i\Gamma_i\delta_{ij}\delta(t-t')$. Parameters $V_{ij}$ are the controlled parameters of the oscillator coupling, with $V_{ij}$ = $V_{ji}$. In the experiments on the effect of the coupling, $D_i$ determines the effective temperature of the noise. We set $D_1 = D_2$. 

We use resonant modulation, $|\omega_p-2\omega_i|\ll \omega_i$, which allows us to parametrically excite vibrations even with small $F_p$. In the absence of resonator coupling and noise the two stable vibrational states of $i$th resonator are

\begin{align}
\label{eq:stable_states}
q_i(\sigma_i;t) = A_i\sigma_i\cos[(\omega_p/2)t + \varphi_i],
\end{align}
where $A_i$ and $\varphi_i$ are the vibration amplitude and phase, and $\sigma_i=\pm 1$. The values of $A_i,\varphi_i$ depend on the resonator parameters; for  small damping $|\varphi_i|\ll 1$. For brevity, and where it may not cause confusion, we use $\uparrow$ and $\downarrow$ for $\sigma_i=1$ and $\sigma_i=-1$, respectively.

In what follows we associate the vibrational states (\ref{eq:stable_states}) with spin states. This association is justified provided the change of these states because of coupling the oscillators to each other, i.e., the change of the amplitudes $A_i$ and phases $\varphi_i$, is small. The weakness of the coupling is thus a major condition of the mapping of the system of oscillators on the system of coupled spins.

Classical and quantum noise causes transitions between the states $\sigma_i=\pm 1$ of an isolated oscillator. By symmetry, the rates $W_i(\sigma_i)$ of transitions $\sigma_i\to -\sigma_i$ of the $i$th oscillator are the same for the both states. For weak noise, the transitions are rare, $W_i(\sigma_i)\ll \Gamma_i$, and the dependence of the switching rate on the noise intensity is given by the activation law \cite{Dykman1998,Marthaler2006}. For classical noise

\begin{align}
\label{eq:activation_law}
W_i(\sigma_i) = C_i\exp[-R_i(\sigma_i)/D_i] 
\end{align}
where $R_i(\sigma_i)=R_i(-\sigma_i)$ is the effective activation energy and $C_i\sim \Gamma_i$. Activated switching in single parametric oscillators has been measured in a number of systems \cite{Lapidus1999,Kim2006,Chan2007}. 

In our experiment, the switching rate of each resonator is extracted from the Poisson distribution of the residence times (Appendix B). Due to slight difference in the damping constants, the switching rates for the two resonators are measured to be different. We verify that, when the coupling is zero, in each resonator the two coexisting states with opposite phases are equally occupied and the populations of all 4 states $\sigma_{1,2}=\pm 1$ are equal (Supplementary Note 3).

As seen from Eqs.~(\ref{eq:eom}) and (\ref{eq:stable_states}), if the noise and the coupling of the oscillators are weak,  
to describe the effect of the coupling to the $j$th oscillator on the dynamics of the $i$th oscillator,
one can replace the coordinate of the $j$th oscillator $q_j(t)$ in the equation of motion of the $i$th oscillator (\ref{eq:eom}) by $q_j(\sigma_j;t)$. In this approximation, the $i$th oscillator is driven by a force at frequency $\omega_p/2$ exerted by the oscillators  to which it is coupled. The force changes when the $j$th oscillator switches between its vibrational states. 

The effect of weak coupling can be understood if one considers the dynamics of an isolated parametric oscillator driven by a weak extra force $F_d\cos[(\omega_p/2)t+\phi_d]$ that mimics the force from other oscillators \cite{Dykman2018}. Such force breaks the symmetry of the vibrational states $\sigma_i=\pm 1$. A major consequence of the symmetry lifting for weak force is the change of the switching rates $W(\sigma_i)$. To leading order in $F_d$ this change has been predicted \cite{Ryvkine2006a} to be described by an increment of the activation energy that is linear in $F_d$, 

\begin{align}
\label{eq:log_sus}
R_i(\sigma_i) = \bar R_i + \Delta R_i(\sigma_i),\quad \Delta R_i(\sigma_i) = \chi_i\sigma_i F_d\cos(\phi_d+\delta_i)
\end{align}
Here $\bar R_i$ is the value of $R(\sigma_i)$ in the absence of the drive. The parameters $\chi_i$ and $\delta_i$ are the magnitude and phase of the {\it logarithmic susceptibility}, i.e., the susceptibility of the logarithm of the switching rate. They strongly depend on the parameters of the oscillator and the parametric modulation, but are independent of $F_d$ and $\phi_d$ \cite{Ryvkine2006a}. 

As seen from Eqs.~(\ref{eq:activation_law}) and (\ref{eq:log_sus}),  for small noise intensity even a weak drive can significantly change the switching rates. It therefore can significantly change the stationary populations of the states $w_\mathrm{st}(\sigma_i)$: from the balance equation for these populations $\dot w(\sigma_i) = -W(\sigma_i)w(\sigma_i) + W(-\sigma_i)w(-\sigma_i)$ we obtain $w_\mathrm{st}(\sigma_i) = W(-\sigma_i)/[W(\sigma_i) + W(-\sigma_i)]$. A strong population change that periodically depends on phase $\phi_d$  was indeed seen in  experiments \cite{Mahboob2010}. However, the general effect of the linear dependence of $\log [W(\sigma_i)]$ on the drive amplitude of a periodic force in bistable systems has not been demonstrated other than in simulations \cite{Luchinsky1999b}. This effect may be responsible for the deviation of the escape rate from the expected quadratic dependence on the drive amplitude  in Josephson junctions \cite{Devoret1987}.

\begin{figure}
%for reprint
\includegraphics[scale=0.40]{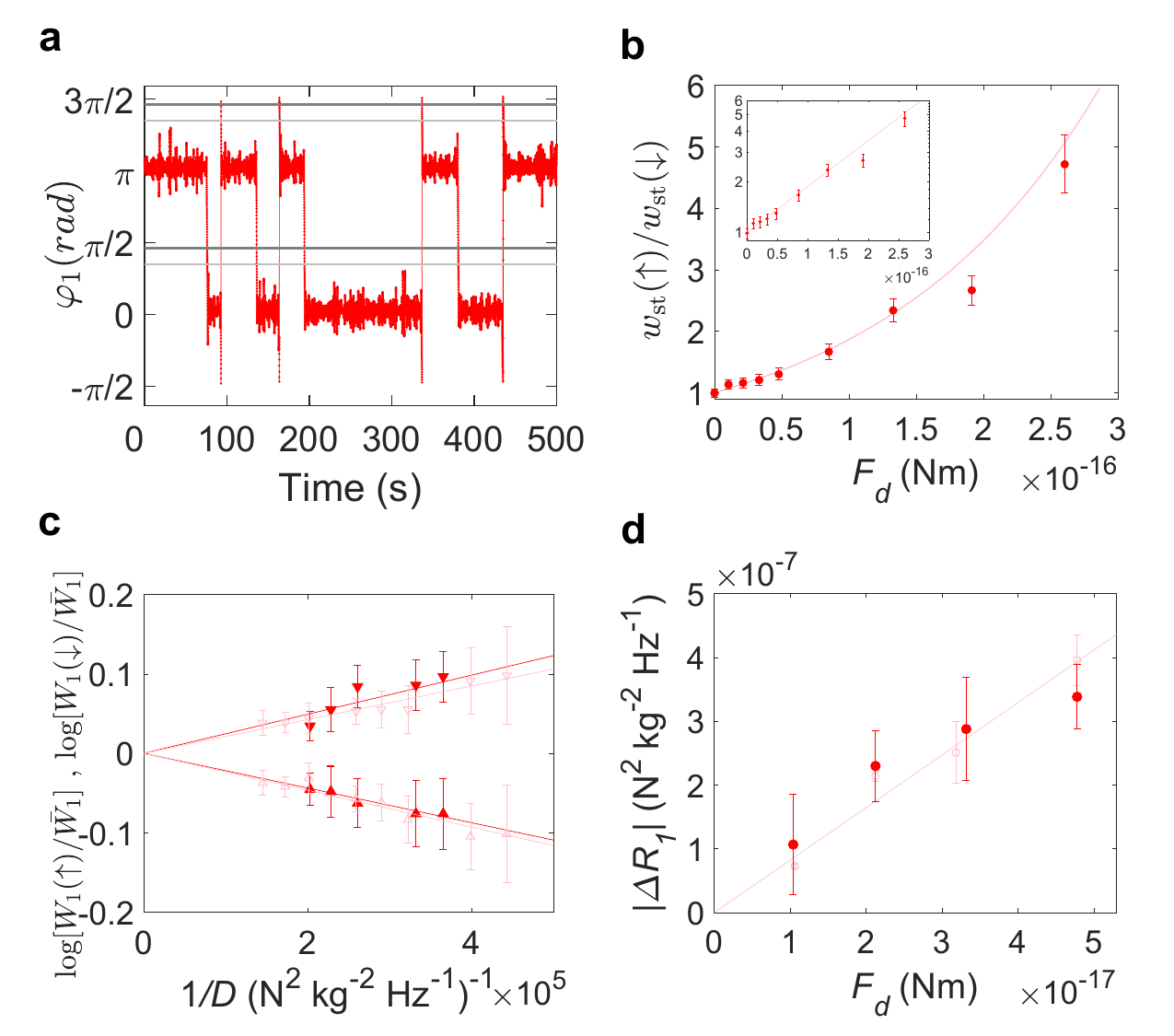}
%for preprint
%\includegraphics[scale=0.8]{Fig2.pdf}
\caption{\textbf{Measurement of the logarithmic susceptibility of a single resonator} Coupling between the two resonators is turned off. We present results for resonator 1 and  indicate the states $\sigma_1=1$ and $\sigma_1=-1$ by $\uparrow$ and $\downarrow$, respectively. \textbf{a.} In the presence of noise, resonator 1 randomly switches between two coexisting vibration states with opposite phase. The two light grey lines are thresholds for identifying phase switches. The dark grey lines represent another choice of threshold. A drive at half the modulation frequency with amplitude $F_{d} = 1.04 \times 10^{-17}$ Nm breaks the symmetry and renders the residence times, and thus the stationary populations of the states $\uparrow$ and $\downarrow$ different. 
\textbf{b.} The ratio $w_\mathrm{st}(\uparrow)/w_\mathrm{st}(\downarrow)$ increases as $F_{d}$ increases. Circles are measured results for the chosen drive phase $\phi_d = 3.3^o$. The solid line represents theory calculated using the simulated logarithmic susceptibility. Inset: same data shown in semilog scale. \textbf{c.} Logarithm of the ratio of 
the switching rates from states $\uparrow$ and $\downarrow$ with the resonant drive turned on, $W_1(\uparrow)$ and $W_1(\downarrow)$ (up and down triangles, respectively), to the rate with no drive $\bar W_1= C_1\exp(-\bar R_1/D)$, plotted as a function of $1/D$. The switching rates are modified by different amounts for the two states according to Eq.~(\ref{eq:log_sus})  The increments of the effective activation energies $\Delta R_1(\uparrow)$ and $\Delta R_1(\downarrow)$ are obtained from the slopes of the linear fits through the origin. \textbf{d.} Increment $|\Delta R_1|$ as a function of $F_d$ for resonator 1. The slope of the linear fit through the origin yields $\chi_1 \cos(\phi_d+\delta_1)$ defined in Eq.~(\ref{eq:log_sus}). Measurements are shown in red. Numerical simulations are shown in pink.}
\label{fig:2} 
\end{figure}

We measure the logarithmic susceptibility of each resonator in our two-resonator system. By setting  $V_{cpl}$ = 0 V we ensure there is no coupling between the two resonators. For each resonator, we apply a resonant drive $F_d\cos[(\omega_p/2)t+\phi_d]$ on top of the parametric modulation at $\omega_p$. The drive phase $\phi_d$ is chosen to be $3.3^\circ$ so that the results can be compared to the case of coupled oscillators when coupling is later re-introduced. Figure 2a shows the random switches of the phase of resonator 1 as a function of time at a constant $F_d$ of $1.04 \times 10^{-17}$ Nm. The ratio of populations $w_\mathrm{st}(\sigma_1 = +1)/w_\mathrm{st}(\sigma_1 = -1)$ $\equiv w_\mathrm{st}(\uparrow)/w_\mathrm{st}(\downarrow)$ is obtained by measuring the residence time in the two states $\sigma_1 = \pm 1$. Figure 2b shows that this ratio deviates from 1 as $F_d$ is increased. 

Next, the switching rates are measured by fitting to the Poisson distribution of the residence times (Appendix B). Figure 2c shows the effect of $1/D$ (which mimics the inverse noise temperature)  on the logarithm of the ratio of switching rates with the symmetry breaking drive turned on and off. The upper and lower branches represent decrease and increase of the activation energy respectively, corresponding to opposite signs of $\sigma_1$ in Eq.~(\ref{eq:log_sus}). We obtain the increment $|\Delta R_1|$  from the average of the magnitude of the slopes of the two linear fits through the origin. The linear dependence of $\log[W_1(\sigma_1)/\bar W_1]$ on $1/D$ in Fig.~2c confirms that the effect of a weak symmetry-breaking drive  is primarily a change $\Delta R_1(\sigma_1)$ of the activation energy of interstate switching. If $D$ is small compared to $|\Delta R_1|$, the change of the switching rate can be substantial. As shown in Fig.~2d, $|\Delta R_1|$ is indeed linear in $F_{d}$ for a weak drive. The factor $\chi_1 \cos(\phi_d+\delta_1)$ for resonator 1 is given by the slope of the linear fit (solid red line). Measurements are then repeated for resonator 2 (Supplementary Note 4) to yield $\chi_2 \cos(\phi_d+\delta_2)$.
 
In Fig.~2d the measurements  are compared with the results of simulations of the switching rate. There is excellent agreement  between measurement and the general expressions (\ref{eq:activation_law})  and (\ref{eq:log_sus}). However, for stronger drive the dependence of $\log [W_i(\sigma_i)]$ on $F_d$ becomes nonlinear (Supplementary Note 6).

\subsubsection{Switching rates in the system of coupled oscillators}

The above results suggest that, if we now consider coupled oscillators, the rate of switching $\sigma_i\to -\sigma_i$ of the $i$th oscillator depends on the states $\{\sigma_j\}$ of the oscillators coupled to it. From Eqs.~(\ref{eq:stable_states}) - (\ref{eq:log_sus}), for weak coupling it has the form 

\begin{align}
\label{eq:full_rates}
&W_i(\sigma_i,\{\sigma_{j\neq i}\}) = \bar W_i \exp[-\sum_{j\neq i}K_{ij}\sigma_i\sigma_j],
\\
\label{eq:J_ij}
&K_{ij} = V_{ij} \chi_i A_j\cos(\phi_j+\delta_i)/D_i,
\end{align}
where $\bar W_i=C_i\exp(-\bar R_i/D_i)$ is the switching rate in the absence of coupling. The change of the activation energy $\Delta R_i(\sigma_i,\{\sigma_{j\neq i}\})$ is equal to $\sum_{j\neq i}K_{ij}\sigma_i\sigma_j D_i.$

Equation (\ref{eq:full_rates}) has the form of the expression for the switching rates of coupled Ising spins. In the standard Ising model $K_{ij}$ is given by the ratio of the coupling energy $J_{ij}$ to $k_BT$ \cite{Glauber1963}. Therefore $K_{ij} = K_{ji}$. In our case, if all oscillators are identical, we also have $K_{ij}=K_{ji}$, as seen from Eq.~(\ref{eq:J_ij}). Therefore the system of coupled identical parametric oscillators maps onto the standard Ising model of coupled spins.

If the oscillators are different, $K_{ij}\neq K_{ji}$. As $V_{ij} = V_{ji}$ in Eq.~(\ref{eq:J_ij}), the difference originates from both the vibration amplitudes and logarithmic susceptabilities. For $K_{ij}\neq K_{ji}$, the system is mapped onto the {\it asymmetric Ising model}. As seen from the known expressions for the vibration amplitudes and phases as well as the logarithmic susceptibilities (cf. \cite{Ryvkine2006a}), the difference between $K_{ij}$ and $K_{ji}$ can be already large if, for example, the oscillator eigenfrequencies are slightly different: $|\omega_i-\omega_j|\ll \omega_i$, but the ratio $|\omega_i-\omega_j|/\Gamma_i$ is not small and, most importantly, the noise intensity is small. 

The stationary probability distribution $w_\mathrm{st}(\{\sigma_n\})$ is generally not known for the asymmetric Ising model. An important feature of the model is the lack of detailed balance (Appendix D). It leads to the onset of a probability current in the stationary state. 
An elementary transition is a flip of a single spin, with the rate that depends on other spins. The current associated with a flip of the $i$th spin for a given configuration of other spins $\{\sigma_{j\neq i}\}$ is, 

\begin{align}
\label{eq:current_one_spin}
&I(\sigma_i,\{\sigma_{j\neq i}\}\to -\sigma_i, \{\sigma_{j\neq i}\}) \nonumber\\
& = w_\mathrm{st}(\sigma_i,\{\sigma_{j\neq i}\}) \, W_i(\sigma_i,\{\sigma_{j\neq i}\})
 \nonumber\\ &
 - w_\mathrm{st}(-\sigma_i,\{\sigma_{j\neq i}\})\,W_i(-\sigma_i,\{\sigma_{j\neq i}\})
 \end{align}
 %
%Here we use the notation $\{\sigma_{j\neq i}\}$ to indicate the state of all spins other than the spin $\sigma_i$.
For symmetric coupling, $K_{ij}=K_{ji}$, the current (\ref{eq:current_one_spin}) is zero (Supplementary Note 7).

\subsubsection{Measurement of asymmetric coupling constant and probability current}

\begin{figure}
%for reprint
\includegraphics[scale=0.4]{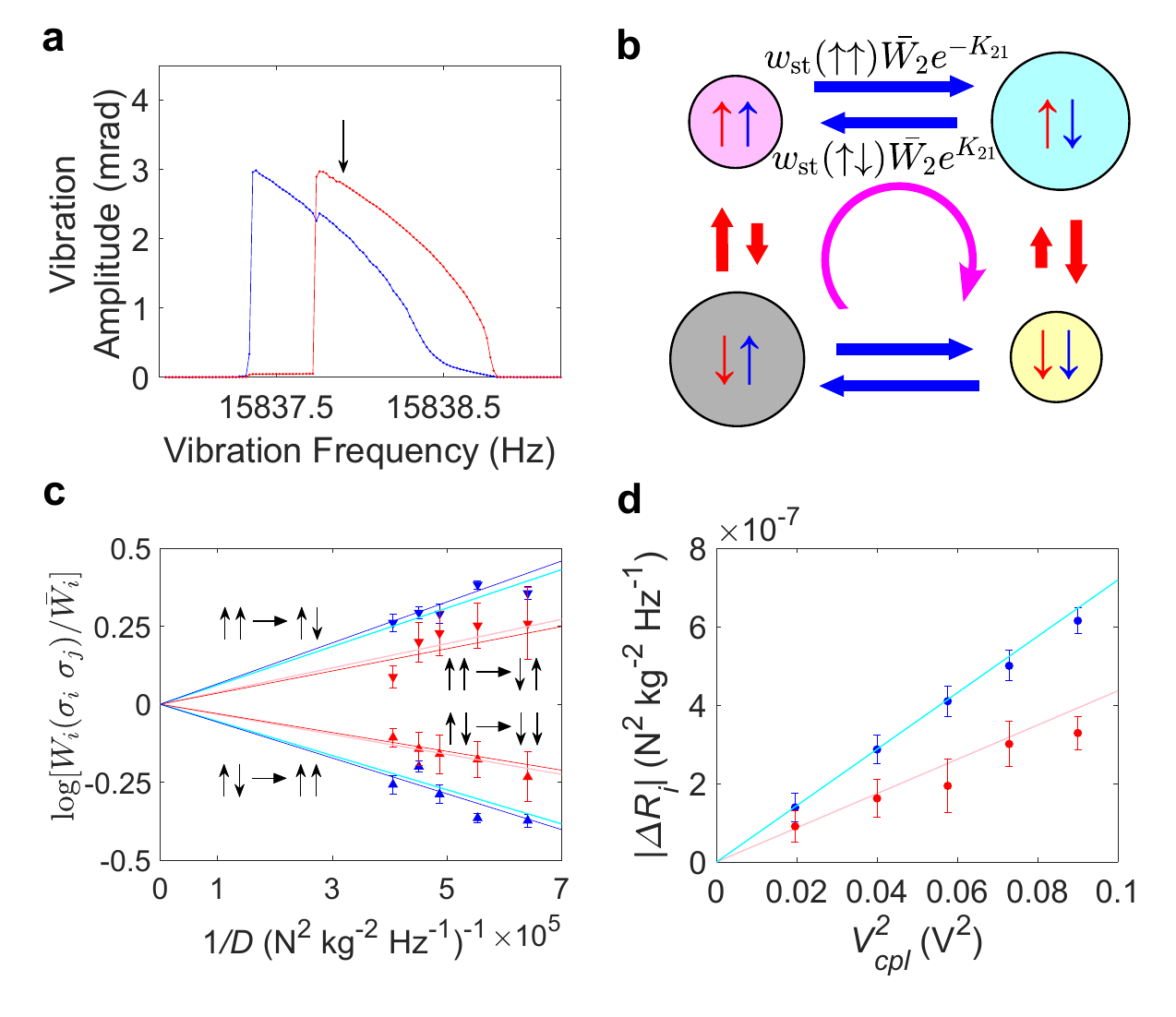}
%for preprint
%\includegraphics[scale=0.8]{Fig3.pdf}
\caption{\textbf{Asymmetric Ising model implemented with two coupled parametric oscillators a.} Vibration amplitudes of resonators 1 (red) and 2 (blue), with $\Delta \omega/2 \pi$ = -0.4 Hz and $V_{cpl}$ = 0.3 V, under identical parametric modulation with no noise added. The arrow marks $\omega_p$/2 for measuring noise-induced switching for the rest of the figure. \textbf{b.} Switchings between the four states of the two-resonator system in the presence of noise. The areas of the circles are proportional to the measured stationary populations $w_\mathrm{st}(\sigma_1,\sigma_2)$ (the first arrow from the left refers to $\sigma_1$ and the second arrow refers to $\sigma_2$). The lengths of the straight arrows between the circles are proportional to the products of the measured switching rates $W_i(\sigma_1,\sigma_2)$ and the corresponding populations $w_\mathrm{st}(\sigma_1,\sigma_2)$. The purple arrow represents the net probability current.  \textbf{c.} Logarithm of the measured changes of the switching rates due to coupling as a function of $1/D$. The values of $\Delta R_i(\sigma_i,\sigma_j) \equiv -D\log[W_i(\sigma_i,\sigma_j)/\bar W_i]$ are determined by the slopes of the linear fits. The difference between $|\Delta R_1(\sigma_1,\sigma_2)|$ and $|\Delta R_2(\sigma_2,\sigma_1)|$ for the same pairs $(\sigma_1,\sigma_2)$ is identified from the different magnitudes of the slopes. This difference determines the asymmetry of the Ising model. \textbf{d.} 
Dependence of $|\Delta R_1|$ (red) and $|\Delta R_2|$ (blue) on $V_{cpl}^2$ that is proportional to the coupling constant. The values of $|\Delta R_i|$ are the average values of  $|\Delta R_i(\sigma_i,\sigma_j)|$ for $\sigma_i=\sigma_j$ and $\sigma_i=-\sigma_j$. The pink and light blue lines are obtained from theory based on the independently simulated logarithmic susceptibilities of individual uncoupled resonators. 
}
\label{fig:3}
\end{figure}

We demonstrate the asymmetry in the coupling coefficients and the existence of a probability current using our system of two coupled parametric oscillators ($i$ = 1, 2). Weak coupling between the two resonators is introduced by applying $V_\mathrm{cpl}$ = 0.3 V.  We adjust $\Delta V_\mathrm{R, 1}$ and $\Delta V_\mathrm{R, 2}$ to tune the resonant frequencies to be close but non-identical, with $\omega_1  - \omega_2 = 0.4$ Hz. The two resonators are subjected to parametric modulation of the same amplitude and the same frequency $\omega_p$. As shown in Fig.~3a, when $\omega_p$ is swept up, resonator 2 undergoes a subcritical bifurcation first, followed by resonator 1. The electrostatic coupling between the two plates favors the configuration where the phases of the resonators are opposite to each other. In the absence of injected noise, resonator 1 adopts a vibration phase opposite to resonator 2 as $\omega_p$ is increased. Correlations in the phase were previously observed in two nanomechanical parametric resonators \cite{Karabalin2011} undergoing supercritical bifurcations. Unlike Ref. \cite{Karabalin2011} where the amplitude increases from zero in a continuous fashion, in our measurement the amplitudes jump sharply from zero in a subcritical bifurcation. 

Next, we fix  $\omega_p$ at $2 \omega_2$ and increase the noise intensity while maintaining the same effective temperatures in the two resonators, $D_1 = D_2 =D$. The noise induces switching of each of the two resonators at random times. We measure the time intervals during which the 4 states are occupied, and obtain the stationary probability distributions $w_\mathrm{st}(\sigma_1,\sigma_2)$. For brevity we indicate the states $\sigma=1$ and $\sigma=-1$ by $\uparrow$ and $\downarrow$, respectively, as we also did in  Fig.~\ref{fig:2}. Therefore the 4 states are $\uparrow \uparrow$, $\uparrow \downarrow$, $\downarrow \uparrow$ and $\downarrow \downarrow$. 
The areas of the circles in Fig.~\ref{fig:3}b are proportional to
the measured stationary probability distribution. We find that $w_\mathrm{st}(\uparrow \downarrow)$ and $w_\mathrm{st}(\downarrow \uparrow)$ exceed $w_\mathrm{st}(\uparrow \uparrow)$ and $w_\mathrm{st}(\downarrow \downarrow)$, consistent with notion that the electrostatic coupling favors opposite vibration phases in the two resonators, so that $K_{12}, K_{21} < 0$. The measured $w_\mathrm{st}$ are in good agreement with Eq.~(\ref{eq:two_spins_stationary}).

The change of the state populations comes from the change of the switching rates. From Eq.~(\ref{eq:full_rates}) applied to two resonators, the rate of switching from the state $\sigma_i$ of resonator $i$ is changed by $\exp(-K_{ij})$ if $\sigma_j=\sigma_i$, i.e., the phases of the two resonators are almost equal, and by $\exp(K_{ij})$ if the phases are opposite. 

For two coupled resonators, there are a total of 8 transitions, as illustrated in Fig.~3b. In the experiment, each of the 8 switching rates is individually measured, by fitting to the Poisson distribution of the residence times. Measurements are performed both before and after the coupling is turned on to give $\bar W_i$ and $W_i(\sigma_i,\sigma_j)$ respectively 
[for two resonators, we use the notation $W_i(\sigma_i,\sigma_j)$ rather than $W_i(\sigma_i,\{\sigma_j\})$].
The ratio $W_i(\sigma_i,\sigma_j) / \bar W_i$  represents the modification of the switching rate of resonator $i$ due to coupling. 

Figure 3c plots the logarithm of the ratio $W_i(\sigma_i,\sigma_j) / \bar W_i$ for the two resonators as a function of $1/D$, where red and blue results correspond to switching of resonator 1 and 2 respectively. 
For the upper branches where the phases are identical, the switching rates are increased due to coupling, and vice versa for the lower branches. The lines  are linear fits through the origin from which the change of the activation barriers $\Delta R_i$ can be obtained by taking the negative values of the slopes. 

We observe that, in agreement with Eq.~(\ref{eq:full_rates}),

\begin{align}
\label{eq:rate_parity}
W_i(\sigma_i,\sigma_j) = W_i(-\sigma_i,-\sigma_j)
\end{align}
within the measurement uncertainty. 
%[for two resonators, we use the notation $W_i(\sigma_i,\sigma_j)$ rather than $W_i(\sigma_i,\{\sigma_j\})$]. 
Therefore in Fig.~3c we show the logarithm of the ratio of the average values of $W_i(\sigma_i,\sigma_j)$ and $W_i(-\sigma_i, -\sigma_j)$ to $\bar W_i$.

There is a clear difference  between the measured values of $|\Delta R_1|$ and $|\Delta R_2|$ for the same sets $(\sigma_1,\sigma_2)$. For resonator 1, the  slopes measured in Fig.~3c are $3.6\times10^{-7}$ N$^2$kg$^{-2}$Hz$^{-1}$ and $-3.0\times10^{-7}$ N$^2$kg$^{-2}$Hz$^{-1}$ for $\sigma_1=\sigma_2$ and $\sigma_1=-\sigma_2$, respectively, whereas those for resonator 2 are $6.6\times10^{-7}$ N$^2$kg$^{-2}$Hz$^{-1}$ and $-5.8\times10^{-7}$ N$^2$kg$^{-2}$Hz$^{-1}$ for $\sigma_1=\sigma_2$ and $\sigma_1=-\sigma_2$, respectively. Averaging the magnitude of the slopes $\Delta R_i(\sigma_i,\sigma_j)$ for $\sigma_i=\sigma_j$ and $\sigma_i=-\sigma_j$ yields $|\Delta R_1|$ exceeding $|\Delta R_2|$ by a factor of 1.7. The difference between $|\Delta R_1|$ and $|\Delta R_2|$ demonstrates that our system of two parametric resonators with different resonant frequencies maps onto the asymmetric Ising model. 

Figure 3d shows that $|\Delta R_1|$ and $|\Delta R_2|$ are proportional to the square of potential difference $V_\mathrm{cpl}$ between the two vibrating plates, with different proportionality constants for the two resonators.
The measured values of $|\Delta R_{i}|$ are compared in Fig.~3d with  Eq.~(\ref{eq:J_ij}) evaluated with the numerically simulated values of the logarithmic susceptibility and the vibration amplitudes and phases $A_j, \phi_j$ independently calculated for each resonator in the absence of coupling. There is good agreement between the entirely independent measurements with the coupling (circles) and the simulations with no coupling [the lines based on Eq.~(\ref{eq:J_ij})]. In turn, the simulations with no coupling are in excellent agreement with the measurement of the logarithmic susceptibility with no coupling, as seen from Fig.~\ref{fig:2}. The linear dependence of $\log[W_i(\sigma_i,\sigma_{j}) / \bar W_i]$ on $1/D$ in Fig.~3c confirms the proposed mechanism of the strong effect of even weak coupling, provided the noise is also weak.

As discussed earlier, a difference between $|\Delta R_{1}|$ and $|\Delta R_{2}|$, and hence $K_{12}$ and $K_{21}$, implies that detailed balance is broken, giving rise to a net probability current. 
For two resonators, the stationary probability distribution can be calculated (Appendix C), and then Eq.~(\ref{eq:current_one_spin}) gives

\begin{align}
\label{eq:two_spin_current}
&I(\uparrow\uparrow\,\to \, \uparrow\downarrow) %\nonumber\\
%&
= \frac{\bar W_1 \bar W_2}{2} \frac{\sinh(K_{12}-K_{21})}
{\bar W_1 \cosh(K_{12}) +\bar W_2 \cosh(K_{21})}.
\end{align}
In the experiment, the probability currents are obtained by taking the product of the measured stationary probability distribution and the measured switching rate out of the specific state. They are represented by block arrows in Fig.~3b. The lengths of the arrows are chosen to be proportional to the product of the measured stationary probability distribution and the measured switching rate. Our measurement demonstrates the lack of detailed balance, as evident from the difference in length of each pair of arrows. The magnitude of the net probability current for the four branches are identical within the measurement uncertainty (Supplementary Note 5). As denoted by the purple arrow in Fig.~3b, the net probability current flows in the clockwise direction for $\omega_2 - \omega_1$ = -0.4 Hz.

\begin{figure}
\begin{center}
\includegraphics[scale=0.6]{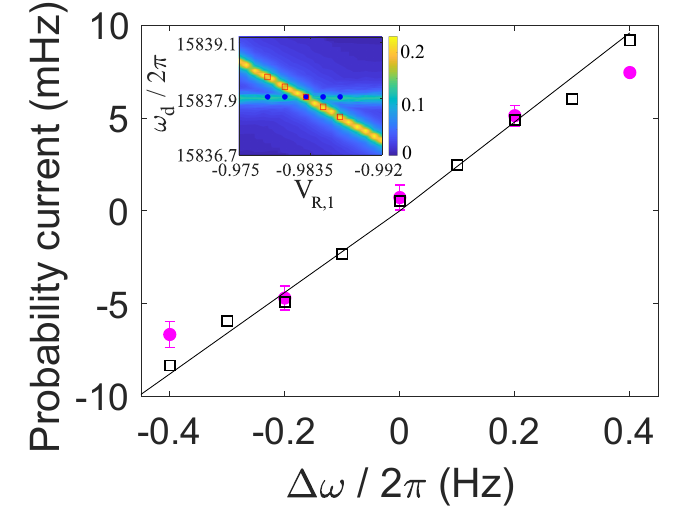}
\end{center}
\caption{ \textbf{Probability current for two non-identical coupled parametric resonators} Dependence of probability current on the frequency mismatch $\Delta \omega$ between the two resonators at $V_\mathrm{cpl}$ = 0.3 V. Purple circles are measurement. Calculations for two resonators based on the simulated logarithmic susceptibility of individual units are plotted in black. The straight line is a linear fit through the origin. Inset: For the considered weak coupling the frequency anticrossing as a function of $\omega_2 - \omega_1$ is undetectable. The color represents the sum of the amplitudes of forced vibrations of the two modes in nm; $x$-axis is the bias $V_{R,1}$ (V) that controls $\omega_1$, whereas $y$-axis is the frequency of resonant drive (Hz) applied to both resonators. Red squares and blue circles mark the values of $\omega_1$ and $\omega_2$ used in the main figure. } 
\label{fig:4}
\end{figure}

In our system of two coupled resonators with near identical damping, the sign and magnitude of the probability current are largely determined by the frequency mismatch $\Delta \omega = \omega_2 - \omega_1$ if the coupling and the noise intensity are fixed. When $\Delta \omega$ is changed to 0.4 Hz by adjusting $V_\mathrm{R,2}$, we find that the sign of the net probability current is reversed. Figure 4a plots the net probability current averaged over the four branches as a function of $\Delta \omega$. The line represents the probability current predicted by Eq.~(\ref{eq:two_spin_current}) with $K_{12}$ and $K_{21}$ given by the simulated value of the logarithmic susceptibilty of a single resonator using Eq.~(\ref{eq:J_ij}).

The difference between $K_{12}$ and $K_{21}$, and hence the probability current, can be tuned to zero by choosing $\Delta \omega$. In our system, choosing $\Delta \omega$ equal to zero makes the probability current vanish within measurement uncertainty. Detail balance is restored. The two resonators therefore map to the symmetric Ising model. The stationary distribution $w_\mathrm{st}$ found in the experiment in this case coincides with the standard expression $w_\mathrm{st}(\{\sigma_i\}) \propto \exp(\sum K_{ij}\sigma_i\sigma_j/2)$ (Supplementary Note 8). We further show in the SM that while $w_\mathrm{st}(\uparrow \downarrow) =w_\mathrm{st}(\downarrow \uparrow)$ exceed $w_\mathrm{st}(\uparrow \uparrow)= w_\mathrm{st}(\downarrow \downarrow)$ due to the coupling, the switching rates given by Eq.~(\ref{eq:current_one_spin}) lead to vanishing of the net probability current.

\section{Discussion}

Our results demonstrate that a system of slightly different parametric oscillators provides a long-sought  inorganic implementation of an asymmetric Ising model. The parameters of the model are determined by the oscillator parameters, including the eigenfrequencies and the coupling, as well as the amplitude and frequency of the parametric modulation. These parameters can be controlled in a broad range. For oscillators based on micro- and nanomechanical resonators, this opens a way of creating asymmetric Ising networks with variable coupling strength and variable connectivity, which is the problem of interest for diverse disciplines, from biology to artificial intelligence. Besides these applications, such networks provide a conceptually simple setting for studying features of many-body dynamics away from thermal equilibrium. One of the major generic features is the lack of detailed balance, which leads to the onset of a probability current in the stationary state. 

We measured the stationary probability current in an asymmetric Ising system. The magnitude of the current depends exponentially strongly on the interrelation between the coupling of the oscillators and the intensity of the noise in the range where both are small. Our analysis and measurement are done in the regime where the coupling-induced change of the oscillator frequencies is much smaller than the frequencies themselves, and the noise-induced spread of the vibration amplitudes is much smaller than the amplitudes themselves. Yet the ratio of the properly scaled coupling and noise intensity can be arbitrary. We note that, for a parametrically excited oscillator, noise necessarily comes along with relaxation, so that it is present even in the quantum regime.

The experiment shows that the effect of weak coupling of parametric oscillators can be quantitatively described in terms of an entirely independent  effect of an additional drive at half the modulation frequency applied to an individual oscillator. It is demonstrated that, in a broad range of the drive amplitudes, the drive leads to a change of  the {\it logarithm} of the  rate of switching between the  vibrational states of the oscillator, which is linear in the drive amplitude. The corresponding logarithmic susceptibility was measured and found to be in excellent agreement with simulations.

The stationary state of an asymmetric Ising model is not known, generally. This is not a consequence of disorder. A simple ``ordered'' system that maps onto an asymmetric Ising system is a chain of parametric oscillators where the coupling to the nearest neighbors for the oscillators on even and odd sites is different. The coupling parameters take on two values, $K_\mathrm{e} = K_{2n\,2n\pm1}$ and $K_\mathrm{o}=K_{2n+1 \, 2n+1\pm 1}$. For small $|K_\mathrm{e,o}|$ one can analyze the spin dynamics similar to how it was done by Glauber \cite{Glauber1963} for a symmetric chain (Supplementary Note 9). In particular, we find that there are two spin-diffusion waves for a periodic chain; in an asymmetric model the wave frequencies, rather than being purely imaginary, can be complex, generally. The probability current in the stationary state is $\propto K_\mathrm{o} - K_\mathrm{e}$. This model immediately extends to a square lattice, which can address the question of the possibility of an Onsager-type transition for an asymmetric Ising model with nearest-neighbor coupling.  We note that, for an asymmetric Ising model, the eigenvalues of the balance equation can be complex in the general case, in contrast to a symmetric Ising model.

\hfill
\section{Acknowledgement}
This work is supported by the Research Grants Council of Hong Kong SAR (Project No.~16304620) and partially supported by Project No. HKUST C6008-20E. MID was supported in part by the National Science Foundation through
Grant No.~DMR-2003815.
%\noindent
%{\bf Methods}
%\appendix
%\small
\noindent
\section*{Appendix A: Excitation and detection schemes.} 
For each resonator $i$ ($i$ = 1, 2) shown in Fig.~1b, the top plate is subjected to electrostatic torques exerted by the left and right electrodes underneath. If the potential difference between the two top plates $V_{cpl} = V^{\mathrm{top}}_1 - V^{\mathrm{top}}_2$ is non-zero, there is also an electrostatic attraction between the two resonators. Each top plate is connected to the input of an amplifier that is a virtual ground for ac voltages. On the left electrodes, the ac component $V_{p,i} \cos(\omega_p t)$ controls the modulation of the spring constant via electrostatic springs softening. When a symmetry breaking torque is needed to measure the logarithmic susceptibility, a second ac component $V_{d,i} \cos[(\omega_p/2) t + \phi_i]$ is added. The noise voltage $V_{n,i}(t)$ generates the noise torque to induce transitions between the two states.

Voltages on the right electrodes only contain dc components. They are adjusted for fine tuning of the resonant frequencies of the two resonators in order to maintain the desirable difference of the oscillator frequencies $\Delta \omega$. Coupling between the two resonators is controlled by the potential difference between the top plates via $V_\mathrm{cpl}$  (Supplementary Note 1).

Vibrations in each resonator are detected by measuring the change of capacitance between the top plate and the two underlying electrodes. The dc voltages described above lead to build up of charges on the top plates. As each of the top plates rotates, the capacitances with the two underlying electrodes change. Charge flowing out of the two top plates are detected independently by two separate amplifiers. The outputs of each amplifier is fed into a lock-in amplifier referenced at $\omega_p/2$.

\hfill

\noindent
\section*{ Appendix B: Measurement of switching rates.}
To measure the switching rate of an individual resonator, its oscillation phase $\varphi$ is recorded as a function of time using a lockin amplifier. Figure 2a shows part of a record for resonator 1. If the resonator initially resides in the state $\sigma = -1$ with $\varphi \approx \pi$, we identify that it has switched to the $\sigma = +1$ state with $\varphi \approx 0$ when the phase  goes over the threshold $\ep$, where $\pi/4 < \ep< \pi/2$. In switching from the initial state $\sigma=+1$ with $\varphi \approx 0$ the phase with overwhelming probability jumps to $-\pi \equiv \pi (\mathrm{mod} 2\pi)$. In this case
the threshold is $-\pi+\ep$. As the resonator switches back and forth between the two states, we record two sequences of residence times for the two states separately. The residence times in each state are plotted as a histogram. A typical histogram is shown in Fig.~5. The exponential decrease in the histogram confirms that the transitions are random and uncorrelated in time. An exponential fit to the histograms yields the switching rate. Fitting to a separate histogram gives the rate of switching from another state. We check that the measured switching rate does not depend on the choice of $\ep$. For example, in Fig.~2a, the dark and light lines indicates two difference choices of threshold $\ep$. They yield measured switching rates that are equal within the error bar of the fitting.

\begin{figure}[h]
%for reprint
\includegraphics[scale=0.40]{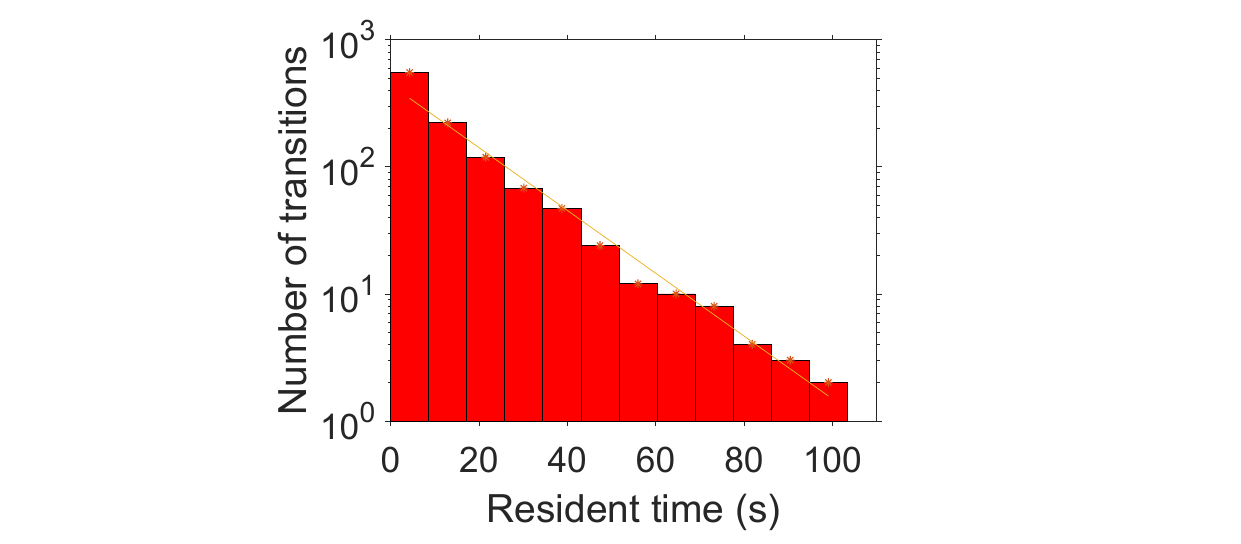}
%for preprint
%\includegraphics[scale=0.8]{Fig5.pdf}
\caption{ \textbf{Histogram of the residence times.} The residence times are recorded for resonator 1 switching out of the $\sigma_1 = + 1$ state at $F_d = 0$, $D$ = $3.01 \times 10^{-6}$ N$^2$ kg$^{-2}$ Hz$^{-1}$ and $\omega_p/2 = \omega_1$. The slope of the linear fit gives the rate of switching out of this state.}
\label{fig:5}
\end{figure}

For uncoupled oscillators in the absence of the symmetry breaking drive, the measured switching rates out of the two states of each resonator are identical to within experimental uncertainty. Their value gives $\bar W_i$ for resonator $i$. Moreover, the stationary probability distributions $w_\mathrm{st}(\uparrow \downarrow)$, $w_\mathrm{st}(\downarrow \uparrow)$, $w_\mathrm{st}(\uparrow \uparrow)$ and $w_\mathrm{st}(\downarrow \downarrow)$ are measured to be equal to within measurement uncertainty (Supplementary Note 3). 

To measure the logarithmic susceptibility of a single resonator, the switching rates are measured after the symmetry breaking drive is turned on. The fractional change of the switching rates for the two states are opposite in sign, as illustrated for resonator 1 in Fig.~2c.  

Logarithmic susceptibility can be calculated using the method of optimal fluctuation \cite{Smelyanskiy1997b} or found from simulations \cite{Luchinsky1999b}. The results have been established to be in excellent agreement. Therefore here we directly used simulations to find the magnitude $\chi$ and the phase $
\delta$ of the logarithmic susceptibility. To do this we incorporated the drive $F_d\cos(\omega_p t/2 +\phi_d)$ into the equation of motion (\ref{eq:eom}) of resonator 1 and set the coupling parameters $V_{ij}$ equal to zero. We then switched to the rotating frame and used the standard rotating wave approximation to reduce the problem to a set of equations for the quadratures of $q_1(t)$. Forced vibrations at frequency $\omega_p/2$ in the lab frame correspond to stable stationary solutions of the equations for the quadratures in the absence of noise. Noise causes switching between these states. The residence times are identified and used to calculate the switching rate in a manner similar to the measurement procedure described above. This allowed us to avoid simulating multiple ($\gtrsim 10^7-10^9$ in our case) oscillations of the parametric oscillator in the lab frame. 

To measure the Ising model parameters $K_{12}$ and $K_{21}$, the switching rates are measured before and after the coupling is turned on. The fractional change of the switching rates for resonators 1 and 2 are plotted in red and blue respectively in Fig.~3c.

\hfill

%\noindent
\section*{Appendix C: The balance equation.} The dynamics of the chain of coupled parametric oscillators is mapped on the dynamics of Ising spins by associating the stable vibrational states of the oscillators with spin-1/2 states. Fluctuations lead to random switching of the spins. The evolution of the distribution $w(\sigma_1,\sigma_2,...)\equiv  w(\{\sigma_i\})$ over the spin states is described by the balance equation, which can be written in the form

\begin{align}
\label{eq:better_Ising}
\dot w(\{\sigma_i\}) = - \sum_i\sigma_i \sum_{\sigma'_i}\sigma'_i\left[W_i(\sigma'_i,\{\sigma_{j\neq i}\})w(\sigma'_i,\{\sigma_{j\neq i}\})  \right]
\end{align}
with the switching rates given by Eq.~(\ref{eq:full_rates}). We note that, even if the rates $\bar W_i$ are different for different spins (different parametric oscillators), but the model is symmetric, $K_{ij} = K_{ji}$, Eq.~(\ref{eq:better_Ising}) has the stationary solution $w_\mathrm{st}(\{\sigma_i\}) = \mathrm{const}\times \exp[\frac{1}{2}\sum_{i,j}K_{ij}\sigma_i\sigma_j]$, which is just the thermal distribution of the conventional symmetric Ising model. 

For a system of $N$ spins (oscillators) Eq.~(\ref{eq:better_Ising}) is a system of $2^N$ equations. For the case of 2 oscillators it can be solved (see Supplementary Note 7 for more details). The stationary probability distribution  is

\begin{align}
\label{eq:two_spins_stationary}
&w_\mathrm{st}(1,1)= w_\mathrm{st}(-1,-1) =\frac{1}{4} 
\frac{\bar W_1 \exp(K_{12}) + \bar W_2 \exp(K_{21})}
{\bar W_1 \cosh(K_{12}) +\bar W_2 \cosh(K_{21})}   ,\nonumber\\   
 &w_\mathrm{st}(1,-1)= w_\mathrm{st}(-1,1) =\frac{1}{4} 
\frac{\bar W_1 \exp(-K_{12}) + \bar W_2 \exp(-K_{21})}
{\bar W_1 \cosh(K_{12}) +\bar W_2 \cosh(K_{21})} 
\end{align}
This expression was used to obtain Eq.~(8) for the probability current and also in Fig.~\ref{fig:3}. For a symmetric system, $K_{12}=K_{21}=K$, we have $w_\mathrm{st}(1,1)/w_\mathrm{st}(1,-1) = \exp(-2K)$ independent of the values of $\bar W_{1,2}$, whereas for an asymmetric system the populations depend on the interrelation between $\bar W_1$ and $\bar W_2$.

\hfill

%\noindent
\section*{Appendix D:  Breaking of the detailed balance.} 
The lack of detailed balance, and thus the onset of the probability current in the asymmetric Ising model can be shown without knowing the stationary distribution. One has to compare the ratio of flipping an $i$th spin back and forth directly or with a $k$th  spin flipped back and forth on the way. For a system with detailed balance the result should be the same. We now compare these ratios. To shorten the notations, we keep in the expressions for the rates only the spins $\sigma_i$ and $\sigma_k$ and explicitly indicate which of them is flipped; other spins are not flipped.  

The detailed balance condition reads

\begin{align}
\label{eq:path_independence}
&\frac{W(\sigma_i,\sigma_k\to -\sigma_i,\sigma_k)}{W(-\sigma_i,\sigma_k\to \sigma_i,\sigma_k)} =\frac{W(\sigma_i,\sigma_k\to \sigma_i,-\sigma_k)}{W(-\sigma_i,\sigma_k\to -\sigma_i,-\sigma_k)}\nonumber\\
&\times
\frac{W(\sigma_i,-\sigma_k\to -\sigma_i,-\sigma_k)}{W(-\sigma_i,-\sigma_k\to \sigma_i,-\sigma_k)}
\times \frac{W(-\sigma_i,-\sigma_k\to -\sigma_i,\sigma_k)}{W(\sigma_i,-\sigma_k\to \sigma_i,\sigma_k)}
\end{align}
For asymmetric Ising model the equality does not hold: the right-hand side has an {\bf extra factor} $\exp\left[4(K_{ik}-K_{ki})\sigma_i\sigma_k\right]$. We note that the result is independent of the switching rates $\bar W_i, \bar W_k$ in the absence of coupling.

%%%%%%%%%%%%%%%%%%%%%%%%%%%%%%%%%%%%%%%%%%%%%%%%%

\hfill

 %%%%%%%%%%%%%%%%%%%%%%%%%%%%%%%%%%%%%%%%%%%
%\section*{References}
%\bibliographystyle{naturemag}

 %\bibliography{Zotero_July23} 

 %%%%%%%%%%%%%%%%%%%%%%%%%%%%%%%%%%%%%%%%%%%%
 %%%%%%%%%%%%%%%%%%%%%%%%%%%%%%%%%%%%%%%%%%%%%%

% \section{Author contributions}
% M.I.D. and H.B.C. conceived the idea of the work. C.H., M.W., B.Z. and Y.Y. performed the device fabrication and experiments. C.H. analyzed the data.  C.H., M.I.D. and H.B.C. co-wrote the paper.

\hfill
 
 %\section{Additional information}
%{ \bf Supplementary information} accompanies this paper at

%\hfill

%{\bf Competing interests:} The authors declare no competing interests

%\hfill

%{\bf Reprints and permissions} information 

\newpage 
 
 \setcounter{figure}{0}
\renewcommand{\figurename}{Fig.}
\renewcommand{\thefigure}{S\arabic{figure}}

\begin{widetext}

\begin{center}{\Large\bf Supplemental Material: \\``Controlled asymmetric Ising model implemented with parametric micromechanical oscillators''}
\end{center}

%\section{\Large Supplemental Material: \\``Controlled asymmetric Ising model implemented with parametric micromechanical oscillators''}

\section*{
Supplementary Note 1: Excitation and detection schemes}
%\hfill \break

%\setlength{\parindent}{2em}
The top plate of each resonator $i$ ($i$ = 1, 2) is subjected to electrostatic torques exerted by the left and right electrodes. If the potential difference between the top plates $V_\mathrm{cpl} = V_2^\mathrm{top}-V_1^\mathrm{top}$ is non-zero, there is also an electrostatic attraction between the two top plates. Each top plate is connected to the input of an amplifier that is a virtual ground for ac voltages.

For resonator i, the total electrostatic torque from the two electrodes is given by the sum
$\tau_i={\frac{1}{2}}\sum_{\alpha=L, R} (dC_{\alpha, i}/d\theta_i)
\\(V_i^\mathrm{top}-V_{\alpha,i})^2$, 
where the subscript $\alpha$ denotes the left (L) or the right (R) electrode, ${C_{L,i}}$ (${C_{R,i}}$) is the capacitance between the top plate and the left (right) electrode. 
${V_{L,i}}$ and ${V_{R,i}}$ are the voltages on the left and right electrodes, respectively. Expanding the torque about the initial angle ${\theta_{0,i}}$  gives:
\begin{align}
\label{eq:torque1}
\tau_i={\frac{1}{2}} \sum_{\alpha} [C_{\alpha, i}'+C_{\alpha, i}''\theta_i +\frac{1}{2} C_{\alpha, i}'''\theta_i^2+\frac{1}{6} C_{\alpha, i}^{IV}\theta_i^3] (V_i^\mathrm{top}-V_{\alpha,i})^2,
\end{align}
where $\theta_i$ is the rotation angle of resonator $i$ measured from $\theta_{i,0}$. $C_{i}', C_{i}'', C_{i}'''$ and $C_{i}^{IV}$ denotes the derivatives of 
$C_i$ with respect to $\theta_i$, evaluated at $\theta_{0,i}$. Higher order terms are neglected. 

$V_{R, i}$ contains only a dc component that is used to adjust $\omega_i$ via the electrostatic spring softening effect. 
$V_{L, i}$ consists of small ac voltages on top of a dc component $V_{L,i}^\mathrm{dc}$. It is used to control the parametric modulation, the symmetry-breaking torque and the noise torque. 
Considering only the dc component and the term responsible for the parametric excitation
\begin{align}
\label{eq:torque2}
V_{L,i}=V_{L,i}^\mathrm{dc}+V_{p,i} \cos(\omega_pt),
\end{align}
where $V_{p,i} << |V_i^\mathrm{top}-V_{L,i}^\mathrm{dc} |$. To induce transitions between the two coexisting vibration states, a noise voltage $V_{n,i}(t)$ is added to $V_{L,i}$. When the logarithmic susceptibility is measured, an additional ac component $V_d \cos(\frac{\omega_p}{2}t)$ is added. Throughout the experiments we use the modulation frequency $\omega_p=2\omega_2$ when measuring noise induced switching in two coupled resonators.

As seen from Eqs.~(\ref{eq:torque1}) and (\ref{eq:torque2}), the time-independent component of $\tau_i$ modifies the system parameters. The term independent of $\theta_i$ shifts the equilibrium position of $\theta_{i,0}$. The term $\frac{1}{2}\sum_{\alpha} C''_{\alpha, i}\theta_i\Delta V_{\alpha, i}^2$, which is linear in $\theta_i$, leads to electrostatic spring softening due to the static potential differences, producing a shift in the resonant frequency. Here $\Delta V_{L,i}=V_i^\mathrm{top}-V_{L,i}^\mathrm{dc}$  and $\Delta V_{R,i}=V_i^\mathrm{top}-V_{R,i}$. The nonlinear 
%in $\theta_i$ 
restoring torques of $\frac{1}{4} \sum_\alpha C_{i}'''\theta_i^2 \Delta V_{\alpha, i}^2$ and $\frac{1}{6} \sum_\alpha C_{i}^{IV}\theta_i^3 \Delta V_{\alpha, i}^2$ modify the Duffing nonlinearity.

The time-dependent components of $\tau_i$ are responsible for the parametric modulation term and noise term in Eq.~(\ref{eq:torque1}) of the main text, with $F_p=C_{L,i}'' \theta_i \Delta V_{L,i} V_{p,i}$ and $\xi_i (t)=\frac{1}{M_i}  C_i' \Delta V_{L,i} V_{n,i} (t)$.  For the symmetry breaking drive, $F_d=C_i' \Delta V_{L,i} V_{d}$.

Throughout the experiment, the dc voltages $V_2^\mathrm{top}$ and $V_{L,2}^\mathrm{dc}$ for resonator 2 are fixed at 0 V and -1.00 V respectively, so that $\Delta V_{L,2}$ is maintained constant at 1.00 V. For resonator 1, $V_{top,1}$ is changed to control the coupling with resonator 2 as explained in more details below. $V_{L,1}^\mathrm{dc}$ is then adjusted to maintain $\Delta V_{L,1}$ constant at -1.28V. $F_p$ is set to be identical for the two resonators, by choosing $ V_{p,1} \Delta V_{L,1} = V_{p,2} \Delta V_{L,2}$.

The voltages $V_{R,1}$ and $V_{R,2}$  applied to the right electrodes are used to control $\omega_1$ and $\omega_2$ respectively, as mentioned earlier. Unlike $V_{L, i}$, $V_{R,i}$ contains no ac components. Initially, $\Delta V_{R,1}$ and $\Delta V_{R,2}$ are chosen to be -0.5 V and 0.5 V respectively to bring $\omega_1$ and $\omega_2$ close to each other. Subsequently, small changes to $\Delta V_{R,1}$ and $\Delta V_{R,2}$ allow the fine tuning of $\Delta \omega$ to the desired value via the electrostatic spring softening effect discussed above. Furthermore, $V_{R,1}$ and $V_{R,2}$ are adjusted regularly to compensate for long term drifts in the resonant frequencies. The adjustment is performed by first setting $V_\mathrm{cpl}$ to 0 V. Subsequently, the vibration amplitude in response to the parametric modulation is measured at a specific $\omega_p$. The changes in amplitude from the value recorded at the beginning of the experiment are used to infer the shifts in the eigenfrequencies. Small changes to $V_{R,1}$ and $V_{R,2}$ are sufficient to bring the vibration amplitudes, and hence the eigenfrequencies, back to the original values for the experiment to continue.

The two plates have identical width of 200 $\mu$m and thickness of 2 $\mu$m. Coupling between them is generated when $V_\mathrm{cpl} =V_1^\mathrm{top}-V_2^\mathrm{top}$ is applied to the interdigitated comb shaped electrodes with separation ranging from 3 $\mu$m to 5 $\mu$m at different locations. Even though the equilibrium positions of $\theta_{1,2}$, as counted from the horizontal axis in Fig.~1b in the main text, are non-zero, the static rotations are small. When there are no vibrations, the sidewalls on the two plates remain largely parallel and aligned with each other. The electrostatic potential energy between the two top plates is given by $\frac{1}{2} C_{12}(\theta_1,\theta_2) V_\mathrm{cpl}^2$, where $C_{12}$ is the capacitance between the two plates. If we disregard the small misalignment of the plates, $C_{12}$ contains a term $\lambda (\theta_1-\theta_2)^2$, where $\lambda$ is a proportionality constant. This term determines the coupling energy in Eq.~(\ref{eq:torque1}) of the main text. Specifically, $V_{12} = V_{21}  = \lambda V_\mathrm{cpl}^2$. As shown in the inset in Fig.~4 of the main text, level anticrossing  does not occur for the small $V_\mathrm{cpl}$ used in this paper. The term $\propto \lambda(\theta_1^2 + \theta_2^2)$ in the coupling energy leads to changes of $\omega_1$ and $\omega_2$, by $\delta\omega_1$ and $\delta\omega_2$ respectively, with $\delta\omega_1 \sim \delta\omega_2$. These changes are measured at the beginning of the experiment from the shifts in the peaks of the linear resonant response from their value at $V_\mathrm{cpl}$ = 0 V. As described earlier, in the procedure to compensate for the long term drifts in $\omega_{1,2}$, $V_\mathrm{cpl}$ is set to 0 V. After the compensation is completed, $V_\mathrm{cpl}$ needs to be changed back to the target value. The previously recorded values of $\delta\omega_{1,2}$ are used to adjust the applied voltages on the electrodes. In particular, $\omega_p$ is changed by 2$\delta\omega_2$ so that the parametric drive frequency coincides with $\omega_2$. $V_{R,1}$ is also adjusted to account for the small difference between $\delta\omega_1$ and $\delta\omega_2$ so that $\Delta\omega \equiv \omega_2- \omega_1$ is maintained at the desired value.

%\hfill

%\newpage
%\setlength{\parindent}{0em}
%%\selectfont
%\noindent
\section*{Supplementary Note 2: Generation of noise}
\setlength{\parindent}{2em}
%\hfill \break

Voltage noise $V_{n, i}(t)$ is applied on the left electrode of resonator $i$ to induce transitions between the two coexisting states of opposite phases. The noise voltage originates from the Johnson noise of a 50 $\Omega$ resistor at room temperature. After amplification, the noise voltage is bandpass filtered with center frequency $f_0$ = 3000 Hz and bandwidth $f_\mathrm{bd} = 40$Hz. For resonator $i$, the filtered noise voltage is then mixed with a carrier voltage at frequency $f_{c,i}$ to generate two sidebands centered at $f_{c,i} \pm f_0$. For instance, in Fig.~2 of the main text, $f_{c1}$ and $f_{c2}$ are 12841.8 Hz and 12851.8 Hz respectively. The resonant frequencies $\omega_{1,2}/2\pi$ lie within the corresponding upper sideband. The frequency difference $f_{c,2} - f_{c,1}$ is chosen to be much larger than the frequencies $\Gamma_i/2 \pi$, $3\gamma_i A_i^2 /2\pi \omega_p$  ($i=1,2$) that characterize the motion of the modes in the rotating frame (here $A_i$ is the vibration amplitude), so that the two resonators are effectively subjected to independent noise voltages, because the relevant spectral components originate from different frequencies of the pass band (note that the above characteristic frequencies are much smaller than 40 Hz). Finally, the noise voltages are multiplied by a factor $c_i$ proportional to $\sqrt{\Gamma_i}⁄\Delta V_{L,i}$  to give $V_{n,i} (t)$ so that the effective temperature is identical in the two resonators. In Eq.~(1) of the main text, $\xi_i (t)=  \frac{1}{M_i}  C_i' \Delta V_{L,i} V_{n,i} (t)$ The noise correlation time is $\sim 2\pi/f_\mathrm{bd}$.  Therefore on the time scale of slow motion in the rotating frame the noise is effectively $\delta$-correlated.
%\newline

%\setlength{\parindent}{0em}
%\selectfont
\section*{Supplementary Note 3: Fluctuations of the uncoupled resonators}
\setlength{\parindent}{2em}
%\hfill \break

For two uncoupled resonators ($V_\mathrm{cpl} = 0 $ V), the stationary state populations $w_{st}(\uparrow\uparrow), w_{st}(\downarrow\downarrow), w_{st}(\uparrow\downarrow)$ and $w_{st}(\downarrow\uparrow)$ are measured to be near identical (Fig.~S1a). As shown in Fig.~S1b, the rates of switching for resonator 1, $\bar W_1(\sigma_1,\sigma_2 \rightarrow -\sigma_1,\sigma_2)$, are identical for all initial states to within measurement uncertainty. The measured switching rates for resonator 2, $\bar W_2 (\sigma_1,\sigma_2 \rightarrow \sigma_1,-\sigma_2 )$, are also identical for all states. However, the switching rates of different resonators are different: $\bar W_2 (\sigma_1,\sigma_2 \rightarrow \sigma_1,-\sigma_2 )$ exceeds $\bar W_1(\sigma_1,\sigma_2\rightarrow -\sigma_1,\sigma_2)$  by a factor of $\sim 2$ because the resonators have slightly different damping constants.
\begin{figure}[h]
%for reprint
%%\includegraphics[scale=0.57]{FigS1.pdf}
%for preprint
\includegraphics[scale=0.7]{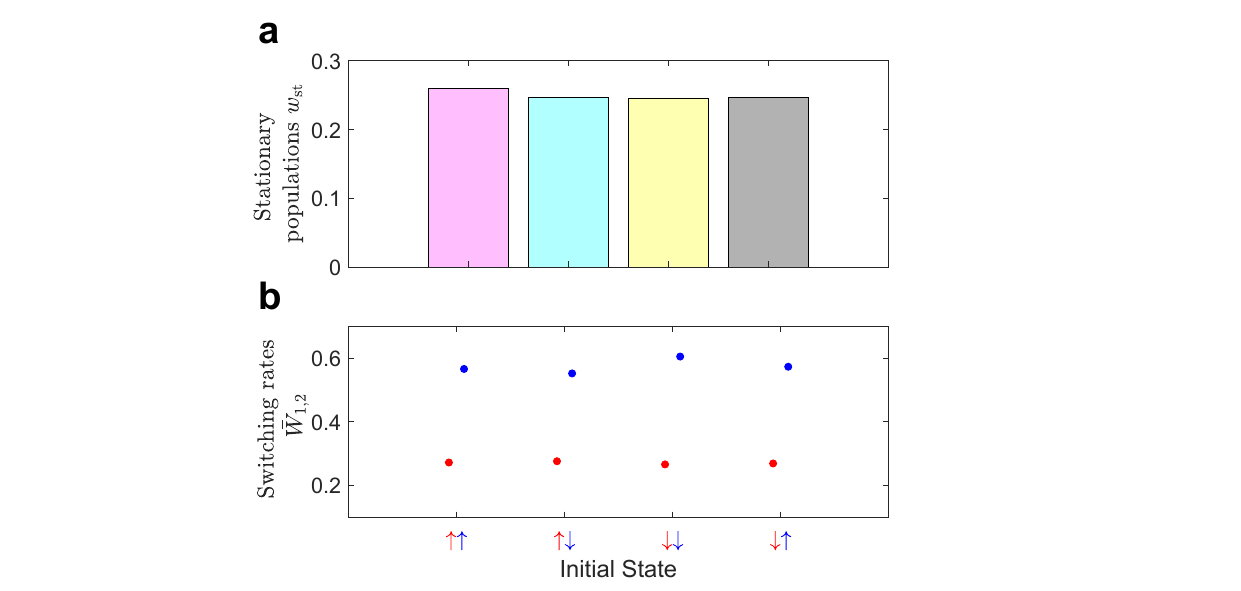}
\caption{\textbf{a.} Stationary populations of the 4 states of two uncoupled resonators.  \textbf{b.} Measured rates of switching out of the states. For the red(blue) circles, only resonator 1 (2) switches. }
\label{fig: S1} 
\end{figure}

%\newpage
% \hfill \break
%\setlength{\parindent}{0em}
%\selectfont
\section*{Supplementary Note 4: Logarithmic Susceptibility of Resonator 2 }
%\setlength{\parindent}{2em}
%\hfill \break

The theory curves for coupled resonators are generated using the simulated logarithmic susceptibilities of individual resonators. Figure 2 of the main text shows how the results of the measurements and simulations of the logarithmic susceptibility of resonator 1. Calculation of the logarithmic susceptibility of resonator 2 follows a similar procedure. First, the changes of the activation barrier are determined as a function of $1/D$ in a fashion similar to Fig.~2c of the main text. Figure S2a shows the ratio of switching rates with and without the symmetry breaking drive for the two states of resonator 2 as a function of $1/D$ at $\omega_p = 2\omega_2$. The slope of the linear fit yields the change of the activation barrier due to the symmetry breaking drive. Figure S2b shows the dependence of the change in activation barrier on $F_d$. The slope of the linear fit gives the logarithmic susceptibility of resonator 2.

\begin{figure}[h!]
%for reprint
%%\includegraphics[scale=0.57]{FigS2.pdf}
%for preprint
\includegraphics[scale=0.7]{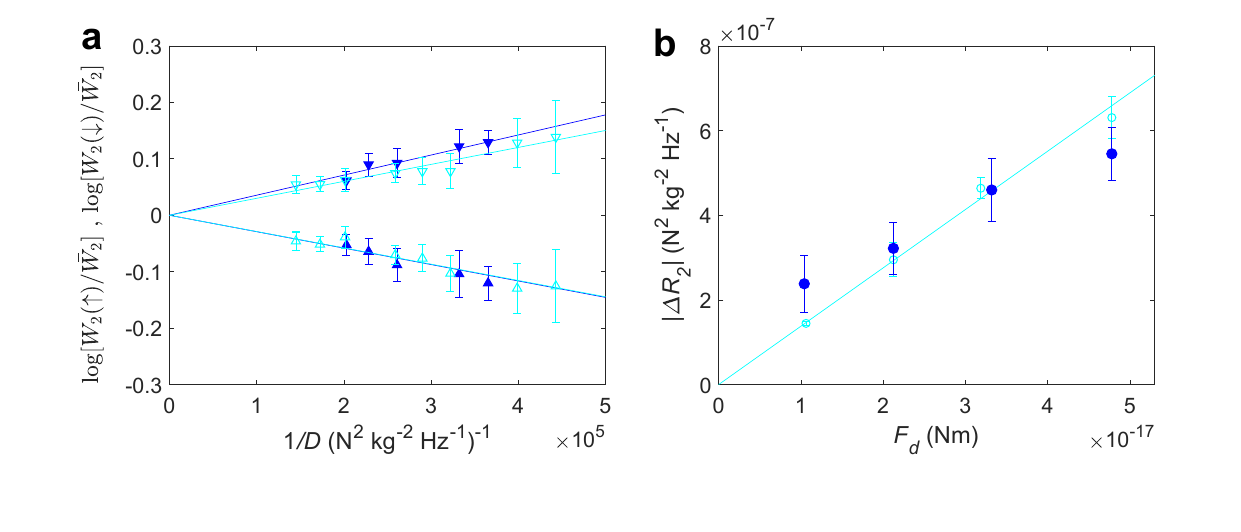}
\caption{\textbf{a.} Dependence of the logarithm of the ratio of switching rates with and without the symmetry breaking drive on $1/D$ at $\omega_p = 2\omega_2$, $F_d = 2.12 \times 10^{-17}$Nm and $\varphi_d = 3.3^o$ for the two states in resonator 2. Up and down triangles represent initial states of $\sigma_2 =$ +1 and -1 respectively.  The straight lines are linear fits through the origin. The negative values of their slopes give the change in activation barrier. \textbf{b.} The change in activation barrier |$\Delta R_2|$ is plotted vs. $F_d$. The line is a linear fit through the origin. The slope of $|\Delta R_2|$ gives the logarithmic susceptibility for resonator 2. Measurements are shown in blue. Numerical simulations are shown in light blue. }
\label{fig: S2} 
\end{figure}

%\hfill \break
%\setlength{\parindent}{0em}
%\selectfont
\section*{Supplementary Note 5: Measurement of probability current}
%\setlength{\parindent}{2em}
%\hfill \break

For two coupled resonators, there are 8 transitions as shown in Fig.~3b of the main text. In the experiment, we measure the switching rate for each transition as well as the stationary populations of the four states. The probability current from one state A to another state B is obtained using the following procedure. First, we evaluate the product of the measured stationary population of state A and the measured transition rate out of state A to state B. Then, we calculate the product of the measured stationary population of state B and the measured transition rate out of state B to state A. In general, these two products are not equal to each other. The net probability current from state A to state B is the difference between these two products. We find that the net probability currents for the four pairs of states are identical to within measurement uncertainty, as shown in Fig.~S3 for three different values of $\Delta \omega$. The values of the probability current plotted in Fig.~4  in the main text represents the mean value of the measured probability current for the four pairs of states in Fig.~S3 for $\Delta \omega= -0.4$ Hz.

\begin{figure}[h!]
%for reprint
%%\includegraphics[scale=0.57]{FigS3.pdf}
%for preprint
\includegraphics[scale=0.7]{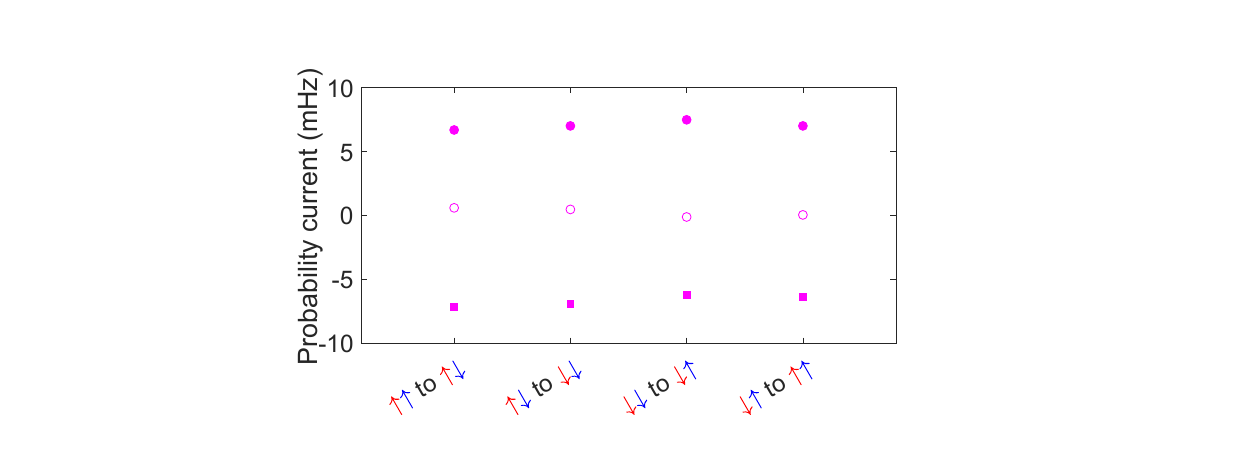}
\caption{\textbf{} The net probability current for each of the four pairs of transitions in the opposite directions for two coupled resonators for $\Delta \omega$ = -0.4 Hz (solid squares), 0 Hz (hollow circles) and 0.4 Hz (solid circles). }
\label{fig: S3} 
\end{figure}

%\newpage
% \hfill \break
%\setlength{\parindent}{0em}
%\selectfont
\section*{Supplementary Note 6: Changes of activation barrier beyond the linear regime}
%\setlength{\parindent}{2em}
%\hfill \break

The effect of the coupling on the switching rate was seen in Ref.~\cite{Alvarez2023}, but the logarithmic-susceptibility regime was not identified there. We emphasize that the mapping on the asymmetric Ising model in our study applies in the regime where the change of the activation barrier has the form $\Delta R_i (\sigma_i,\{ \sigma_{j\neq i} \})=D_i \sigma_{i} \sum_{j \neq i} K_{ij} \sigma_j $. In particular, for two oscillators $ \Delta R_i (\sigma_i,\sigma_j)=-\Delta R_i (-\sigma_i,\sigma_j)=-\Delta R_i (\sigma_i,-\sigma_j)$. The value of $\Delta R_i (\sigma_i, \sigma_j)$ in this regime is determined by the logarithmic susceptibility, and in our experiment we have measured it independently.

The logarithmic-susceptibility regime refers to weak coupling. If the coupling is stronger, even where it does not lead to the onset of new vibrational states, not only are the switching rates modified, but the very stable states of individual oscillators, i.e., their amplitudes and phases, significantly change depending on the states of other oscillators. Therefore one cannot think of an oscillator as having just two states, which underlies the mapping on a spin system.

\begin{figure}[h!]
%for reprint
%%\includegraphics[scale=0.57]{FigS4.pdf}
%for preprint
\includegraphics[scale=0.7]{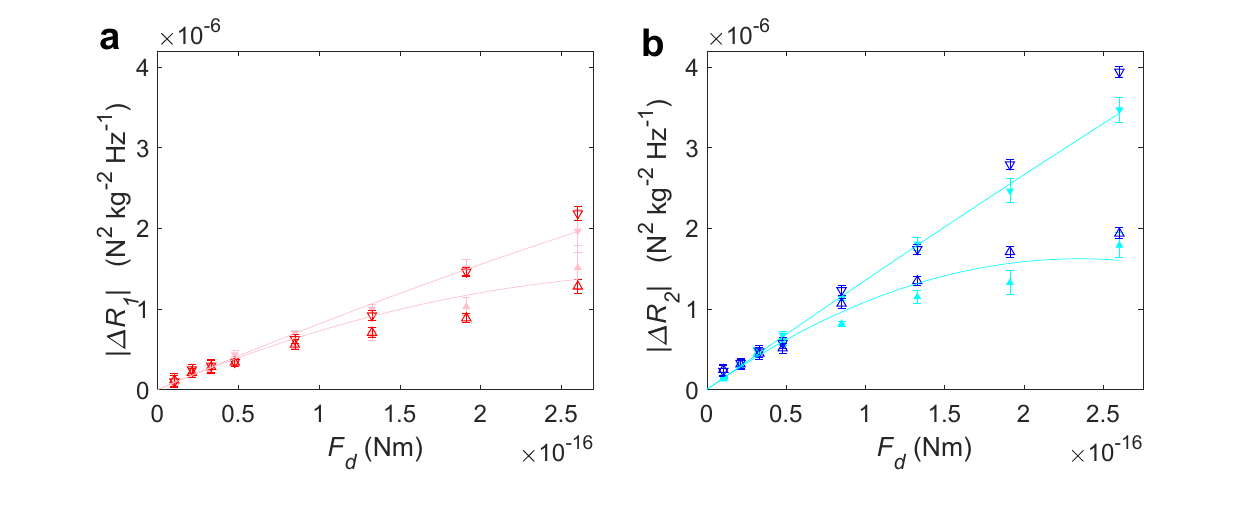}
\caption{\textbf{a.} Increments $|\Delta R_1(\sigma_1=+1)|$ (up triangles) and $|\Delta R_1(\sigma_1=-1)|$ (down triangles) of the activation energy as a function of $F_d$ for resonator 1. The range of $F_d$ is about 5 times larger than in Fig.~2d in the main text. Measurements and numerical simulations are shown in red and pink respectively. The lines are fits to the simulated results by parabolae. They have the same linear term as the line in Fig.~2d. Different quadratic terms are used for $\sigma_1= \pm 1$. \textbf{b.} Similar plot for resonator 2. Measurements and numerical simulations are shown in dark blue and light blue respectively.}
\label{fig: S4} 
\end{figure}

When the amplitude $F_d$ of the symmetry-breaking drive is small for an individual, uncoupled resonator, the change in the vibration amplitudes of the two states is also small. The changes of the activation barriers $\Delta R_i(\sigma_i=+1)$ and $\Delta R_i(\sigma_i=-1)$ are of opposite signs but near identical magnitude. $|\Delta R_i|$ is proportional to the amplitude of the symmetry breaking drive, with a proportionality constant given by the logarithmic susceptibility. As $F_d$ increases, the difference in the vibration amplitudes and phases of the two states cannot be ignored. Figure S4a extends the range of $F_d$ of Fig.~2d in the main text to show that, for resonator 1, $|\Delta R_1(\sigma_1=+1)|$ and $|\Delta R_1(\sigma_1=-1)|$ are no longer equal for large $F_d$. Furthermore, the dependence of $\Delta R_1 (\sigma_1)$ on $F_d$ does not follow a linear relationship. Figure S4b shows a similar plot for resonator 2.

\begin{figure}[h!]
%for reprint
%%\includegraphics[scale=0.57]{FigS5.pdf}
%for preprint
\includegraphics[scale=0.7]{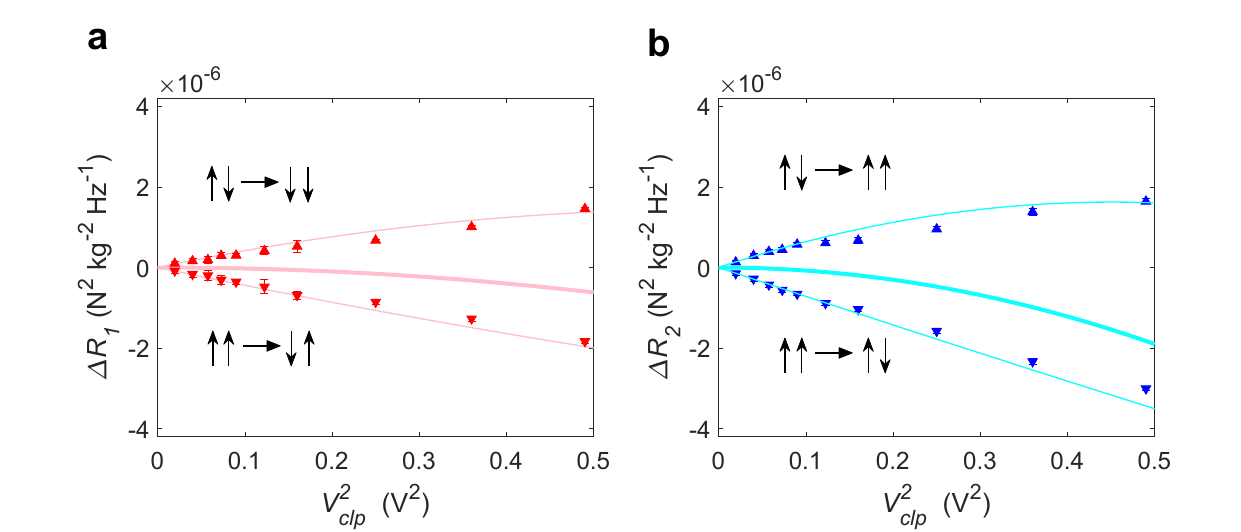}
\caption{\textbf{a. }Dependence of the change $\Delta R_1$ of activation barriers for switching of resonator 1 on $V_\mathrm{cpl}^2$. Initial states with the phase of resonator 1 identical (opposite) to resonator 2 are represented by up triangles (down triangles). Measurements are shown in red. The thin pink lines are obtained from theory based on the effect of a symmetry breaking drive on individual uncoupled resonators using numerical simulations from Figures S4. The thick pink line is a sum of the thin lines. \textbf{b.} Similar plot for resonator 2. Measurements and numerical simulations are shown in blue and light blue respectively.}
\label{fig: S5} 
\end{figure}

% \hfill \break
As described in the main text, when weak coupling is turned on between two resonators, the effects on resonator 1 from the coupling to resonator 2 can be understood in terms of the logarithmic susceptibility of resonator 1 subjected to a symmetry breaking drive, and vice versa for resonator 2. Beyond the weak coupling regime, the increase in switching rates for initial states with identical phases is no longer equal in magnitude to the decrease of switching rates for initial states with opposite phases. In fact, the phase difference between the two vibration states in each oscillator deviates considerably from $\pi$. Figure S5 shows the measured and simulated results for the change of the activation barrier when the range of  $V_\mathrm{cpl}$ is extended beyond that in Fig.~3d in the main text. While Fig.~3d in the main text plots |$\Delta R_{1,2}|$, Figure S5a and Figure S5b show $\Delta R_1(\sigma_1,\sigma_2)$ and $\Delta R_2(\sigma_1,\sigma_2)$ separately and without the absolute value. Since $W_i(\sigma_i,\sigma_j)$ = $W_i(-\sigma_i,-\sigma_j)$ as described by Eq.~(8) in the main text, it follows that $\Delta R_i(\sigma_i,\sigma_j)$ = $\Delta R_i(-\sigma_i,-\sigma_j)$. In Fig.~S5, the up triangles represent the average of the measured values of $\Delta R_i(\uparrow\downarrow)$ and $\Delta R_i(\downarrow\uparrow)$. The down triangles represent the average of $\Delta R_i(\uparrow\uparrow)$ and $\Delta R_i(\downarrow\downarrow)$, both of which are negative. The results for Fig.~3d in the main text are obtained by averaging  $|\Delta R_i(\uparrow\downarrow)|$, $|\Delta R_i(\downarrow\uparrow)|$, $|\Delta R_i(\uparrow\uparrow)|$ and $|\Delta R_i(\downarrow\downarrow)$| for $V_\mathrm{cpl}^{2} < 0.1 $ V$^2$. For $V_\mathrm{cpl}^{2} > 0.1$ V$^2$, the difference in the magnitude of $\Delta R_{1,2}$ for initial states of the same and opposite phases becomes more apparent. In Fig.~S5a and Fig.~S5b, the sum of the top and bottom branches are shown as the thick lines, the deviation of which from zero increases with $V_\mathrm{cpl}^2$.

%\newpage
%\setlength{\parindent}{0em}
%\selectfont
\section*{Supplementary Note 7: Population and transition rates of coupled identical resonators}
%\setlength{\parindent}{2em}
%\hfill \break
In this section, we tune the difference between $K_{12}$ and $K_{21}$ to near zero by choosing $\Delta \omega/2\pi$ = 0 Hz. As a result, the probability current vanishes to within experimental uncertainty. Detailed balance is restored and our system maps onto the symmetric Ising model. 

\begin{figure}[h!]
%for reprint
%%\includegraphics[scale=0.57]{FigS6.pdf}
%for preprint
\includegraphics[scale=0.7]{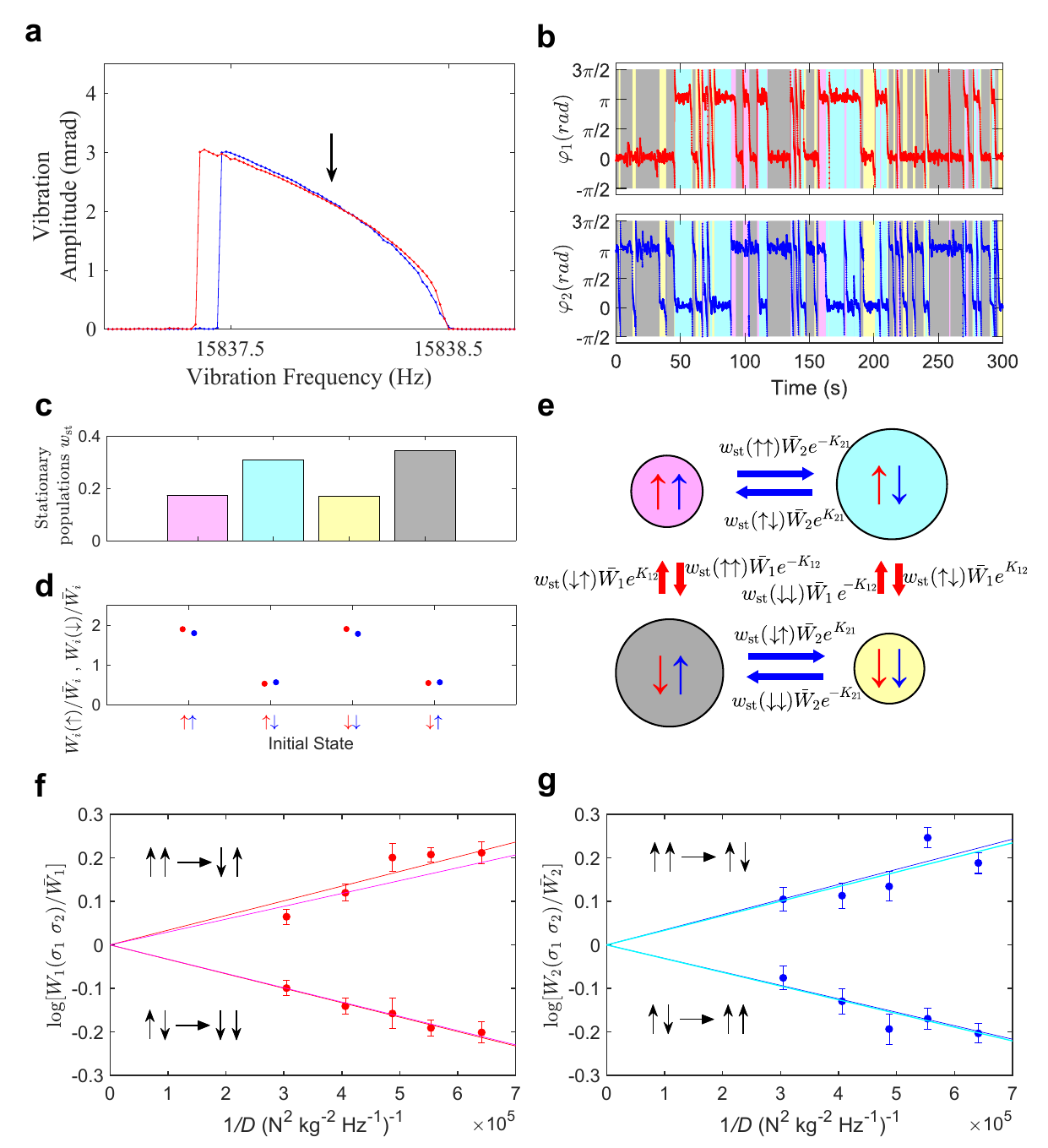}
\caption{\textbf{a. }Vibration amplitudes of resonator 1 (red) and 2 (blue), with $\Delta \omega/2\pi$ = 0 Hz and $V_\mathrm{cpl}$ = 0.3 V, under identical parametric modulation with no noise added. \textbf{b.} Typical records of the phase of the two resonators as a function of time when noise is added. \textbf{c.} Stationary probability distributions of the four states. \textbf{d.} Change of the transition rates from the four initial states due to coupling. For red (blue) circles, only resonator 1 (2) switches. \textbf{e.} Switchings between the 4 states. The areas of circles are proportional to the measured stationary probability distributions $w_{st}(\sigma_1, \sigma_2)$. The lengths of the arrows are proportional to the product of the measured switching rates and the corresponding initial stationary probability distribution. \textbf{f.} Logarithm of the measured changes of the transition rates due to coupling as a function of $1/D$ for resonator 1. The magnitude of the slopes of the linear fits yields $|\Delta R_1|$. \textbf{g.}  Similar plot for resonator 2, with $|\Delta R_2|$ given by the magnitude of the slope of the linear fit. }
\label{fig: S6} 
\end{figure}

Figure S6a shows the vibration amplitudes of the two resonators with $\Delta \omega/2\pi$ = 0 Hz under parametric modulation in the absence of injected noise. The slight difference between the response curves is due to the difference in the decay rates of the modes. Application of noise leads to switching in both resonators, as illustrated in Fig.~S4b. As a result of the coupling ($V_\mathrm{cpl}$ = 0.3 V) that favors opposite phases in the two resonators, the transition rates $W_i (\sigma_i,\{ \sigma_j=\sigma_i \})$ and $W_i (\sigma_i,\{ \sigma_j=-\sigma_i \})$ are increased and decreased compared to the uncoupled values $\bar W_i $ respectively. Moreover, the fractional changes for resonators 1 and 2 are identical to within experimental uncertainty, as shown by the adjacent red and blue circles in Fig.~S6d. Figures S6f and S6g plot the change in transition rates due to the coupling as a function of $1/D$ for resonators 1 and 2 respectively. The magnitudes of the slopes of the linear fits are identical to within $\sim 10 \%$ consistent with $|\Delta R_1| = |\Delta R_2|$ and hence $K_{12}=K_{21}=K$.

We examine how detailed balance is manifested in the probability current, taking the transitions between the $\uparrow\uparrow$ state and $\uparrow\downarrow$ state as an example. The transition rates $W_1 (\uparrow,\uparrow)$ and $W_1 (\uparrow,\downarrow)$ are changed from $\bar W_1$ by factors $\exp\left( -K \right)$ and $\exp (K)$ respectively. The stationary probability distribution for $K_{ij}=K_{ji}$ is $w_{st}=Z^{-1} \exp \left[\frac{1}{2}\sum_{i,j(i\neq j)} K \sigma_i \sigma_j \right]$ where Z is the normalization factor. In the case of two resonators we see that $w_{st} (\uparrow\uparrow)$ and $w_{st} (\uparrow\downarrow)$ are changed from their value $\frac{1}{4}$ in the absence of coupling by $Z^{-1} \exp (K)$ and $Z^{-1} \exp (-K)$ respectively, where $ Z=4 \cosh K$ (Fig.~S6c). The products $W_1 (\uparrow,\uparrow)w_{st} (\uparrow\uparrow)$ and $W_1 (\uparrow,\downarrow)w_{st} (\uparrow\downarrow)$ remain equal in magnitude, as represented by the same length of the two block arrows in Fig.~S6e. The net probability current between the states $\uparrow\uparrow$ and $\uparrow\downarrow$ is therefore zero.

%%%%%%%%%%%%%%%%%%%%%%%%%%%%%%%%%%%%%%%%%%%%%%%%%%%%%%%%%%%%%

%\newpage
%\section*{Balance equation for the state populations}
%\label{sec:balance_eq}
%\setlength{\parindent}{0em}
%\selectfont
\section*{Supplementary Note 8: Balance equation for the state populations}
%\setlength{\parindent}{2em}
%\hfill \break

As discussed in the main text, on the time scale long compared to the relaxation times of the modes, the dynamics of the chain of coupled parametric oscillators is mapped on the dynamics of Ising spins, which switch at random between their states. Two states of an individual spin $\sigma_i=\pm 1$ correspond to two values of the phase of parametrically excited vibrations. The switching rate $W_i(\sigma_i, \{\sigma_{j\neq i}\})$ of the transition $\sigma_i\to -\sigma_i$ of spin $i$ depends on the state of spins $j\neq i$ with which the spin is coupled. The spins are switching at random, and the evolution of the distribution $w(\sigma_1,\sigma_2,...)\equiv  w(\{\sigma_i\})$ over the spin states is described by the balance equation, which can be written in the form
\begin{align}
\label{eq:better_Ising_S}
\dot w(\{\sigma_i\}) = - \sum_i\sigma_i \sum_{\sigma'_i}\sigma'_i\left[W_i(\sigma'_i,\{\sigma_{j\neq i}\})w(\sigma'_i,\{\sigma_{j\neq i}\})  \right].
\end{align}
The switching rates have the form
\begin{align}
\label{eq:rates_from_main}
W_i(\sigma_i,\{\sigma_{j\neq i}\}) = \bar W_i\exp[-\sigma_i\sum_{j}K_{ij}\sigma_j],
\end{align}
and for the spins that model differing oscillators $K_{ij} \neq K_{ji}$, generally, cf. Eqs.~(5) and (6) of the main text.

The evolution of the distribution $w(\{\sigma_i\})$ is determined by the eigenvalues of the $2^N\times 2^N$ matrix $
W_i(\sigma_i,\{\sigma_{j\neq i}\})$, where $N$ is the number of spins. For a symmetric Ising model all eigenvalues of this matrix are real. To see this, we change from the distribution $w$ to $\tilde w$, 
\[w(\{\sigma_i\}) = \exp\left(\frac{1}{4}\sum_{i,j}K_{ij}\sigma_i\sigma_j\right)\tilde w(\{\sigma_i\}).\]
The equation for the function $\tilde w$ takes the form
\begin{align}
\label{eq:master_eq_det_bal}
&\frac{d}{dt}\tilde w(\{\sigma_i) = -\sum_i W_i(\sigma_i,\{\sigma_{j\neq i}\})\,\tilde w(\{\sigma_i\})+ \sum_i \bar W_i \,\tilde w(-\sigma_i, \{\sigma_{j\neq i}\})
\end{align}
The off-diagonal elements of the matrix in the right-hand side,  $\tilde W_i$, are independent of the spin state and are thus the same for $\sigma_i=1$ and $\sigma_i=-1$. Therefore all eigenvalues of Eq.~(\ref{eq:master_eq_det_bal}), and thus also of Eq.~(\ref{eq:better_Ising_S}),  are real. 

For an asymmetric Ising model the eigenvalues of the matrix of the transition rates can be complex. This is another significant difference between symmetric and asymmetric Ising models.

The stationary distribution $w_\mathrm{st}(\{\sigma_i\})$ is given by the solution of Eq.~(\ref{eq:better_Ising_S}) for $\dot w=0$. Formally, the condition $\dot w=0$ makes Eq.~(\ref{eq:better_Ising_S}) a system of $2^N$ linear equations. Only $2^N -1$ of them are linearly independent, as seen from the condition $\sum_{\{\sigma_i\}} \dot w(\{\sigma_i\})\equiv 0$ that follows from Eq.~(\ref{eq:better_Ising_S}). Therefore they have to be complemented by the normalization condition $\sum_{\{\sigma_i\}} w_\mathrm{st}(\{\sigma_i\}) =1$. Only $2^{N-1}$ populations $w_\mathrm{st}(\{\sigma_i\})$ are linearly independent. Indeed, as seen from Eq.~(\ref{eq:rates_from_main}), the rates $W_i(\sigma_i,\{\sigma_{j\neq i}\})$ do not change if we simultaneously change the signs of all spins. Therefore, for any spin configuration $\{\sigma_i\}$ in a finite system, $w_\mathrm{st}(\{\sigma_i\}) = w_\mathrm{st}(\{-\sigma_i\})$ (at this time we do not consider spontaneous symmetry breaking). 

The distribution $w_\mathrm{st}$ can be easily found if the spin system is symmetric, $K_{ij} = K_{ji}$. From Eq.~(\ref{eq:better_Ising_S}), in this case
\begin{align}
\label{eq:symmetric_stat}
 w_\mathrm{st}^\mathrm{sym} = Z^{-1} \exp\left(\frac{1}{2}\sum_{i,j}K_{ij}\sigma_i\sigma_j\right),
 \end{align}
where $Z$ is the normalization factor, which is essentially given by the standard partition function of the Ising system with the appropriately defined coupling constants scaled by the temperature. The relation (\ref{eq:symmetric_stat}) holds even if the chain is disordered, i.e., the values of the prefactor in the switching rate $\bar W_i$ depend on site $i$. It immediately follows from Eqs.~(\ref{eq:rates_from_main}) and (\ref{eq:symmetric_stat}) that the probability current defined by Eq.~(7) in the main text is equal to zero for a symmetric system.

%%%%%%%%%%%%%%%%%%%%%%%%%%%%%%%%%%%%%%%%%%%%%

\subsection{Two-oscillator case}
\label{subsec:two_oscillators}

The balance equation for the populations can be solved in the case of 2 spins, i.e., two parametric oscillators, studied in the experiment. It splits into two uncoupled systems of equations, one  for $w(1,1)-w(-1,-1)$ and $w(-1,1)-w(1,-1)$, and the other for $w(1,1)+w(-1,-1)$ and $w(-1,1) +w(1,-1)$. One of the eigenvalues of these equations is trivial, $\lambda_1=0$. Other eigenvalues are
\begin{align}
\label{eq:two_spins_eigenvalues}
&\lambda_2 = -2(\bar W_1\cosh K_{12} + \bar W_2 \cosh K_{21}),\nonumber\\
&\lambda_{3,4} = \frac{1}{2}\lambda_2 
\pm \left[\frac{1}{4}\lambda_2^2 - 4 \bar W_1 \bar W_2 \cosh(K_{12} - K_{21}) \right]^{1/2}
\end{align}
We see that Re~$\lambda_{2,3,4}<0$, which indicates that the system approaches a stationary state with increasing time. However, the roots $\lambda_{3,4}$ can be complex; for example, for $\bar W_1 = \bar W_2=\bar W$ and $K_{12} = -K_{21}= K$ we have $\lambda_{3,4} = -2\bar W (\cosh J \pm i \sinh J)$. This case corresponds to the parameters $K_{12}$ and $K_{21}$ having opposite signs. For coupled parametric oscillators $K_{ij}$ and $K_{ji}$ are proportional to the same coupling constant, and therefore they have the same signs, so that the eigenvalues (\ref{eq:two_spins_eigenvalues}) are real.

The explicit solution of the balance equation allow one to calculate the correlation functions of the spins. The simplest nontrivial correlator is just $\braket{\sigma_1\sigma_2}$. From the explicit form of the stationary distribution given by Eq.~(D1) of the main text we find
\begin{align}
\label{eq:simple_correlator}
\braket{\sigma_1\sigma_2} =  2 [w_\mathrm{st}(1,1)- w_\mathrm{st}(1,-1)] = 
\frac{\bar W_1 \sinh K_{12} + \bar W_2 \sinh K_{21}}
{\bar W_1 \cosh K_{12} + \bar W_2 \cosh K_{21}}
\end{align}
For small $|K_{12}|, |K_{21}|$ the correlator (\ref{eq:simple_correlator}) is linear in $K_{12}, K_{21}$ , whereas if $K_{12}, K_{21}$ are large and have the same sign $\braket{\sigma_1\sigma_2} \approx \sgn K_{12}$.

We compared Eq.~(\ref{eq:simple_correlator}) with the experimental data. We found a quantitative agreement as a function of the coupling, with the independently measured parameters. This provides an extra proof of the consistence of the evaluation of the stationary distribution and of the picture of switching oscillators as a whole.

\hfill 

%\setlength{\parindent}{0em}
%\selectfont
\section*{Supplementary Note 9: Weakly asymmetric chain}
%\setlength{\parindent}{2em}
%\hfill \break

To illustrate a nontrivial effect of the asymmetry on the stationary distribution we briefly consider the case where the asymmetry is weak,
\begin{align}
\label{eq:weak_asymmetry}
&W_i(\sigma_i,\{\sigma_{j\neq i}\}) = \bar W \exp\bigl(-\sigma_i\sum_kK_{ij}\sigma_j\bigr), 
%
%\nonumber\\&
\quad K_{ij} = \bar K_{ij} + \delta K_{ij}, \quad \bar K_{ij} = \bar K_{ji}, \quad \delta K_{ij} = -\delta K_{ji},
\end{align}
with $|\delta K_{ij}| \ll |\bar K_{ij}|$. To zeroth order in $\delta K_{ij}$, the stationary distribution $w_\mathrm{st}^{(0)}$is given by Eq.~(\ref{eq:symmetric_stat}) for the distribution for a symmetric chain $w_\mathrm{st}^\mathrm{sym}$,
\[w_\mathrm{st}^{(0)} = Z^{-1} \exp\left(\frac{1}{2}\sum_{i,j}\bar K_{ij}\sigma_i\sigma_j\right),
\]
where, again,  $Z$ is an equivalent of the partition function.

%The expression for the correction to $w_\mathrm{st}$ is cumbersome, and we will consider it for a periodic chain with nearest-neighbor coupling, where $K_{ij}\propto \delta _{|i-j|,1}$. 
To simplify the analysis, we will consider a periodic chain with nearest-neighbor coupling and set $\bar K_{i\,i+1} = \bar K$, i.e., $\bar K_{ij}$  is the same for all spins. Furthermore, we will assume that the rates $\bar W_i$ are the same for all spins,  $\bar W_i = \bar W, \forall i$. We will write the first-order correction to the stationary distribution in the form $w_\mathrm{st}^{(0)}(\{\sigma_i\}) w_\mathrm{st}^{(1)}(\{\sigma_i\})$. This correction comes from the first-order (linear in $\delta K_{ij}$) correction to the switching rates in Eq.~(\ref{eq:better_Ising_S}) and is given by equation
\begin{align}
\label{eq:correction_general}
%\sum_i\sigma_i \sum_{\sigma'_i}\sigma'_i W_i^{(0)}(\sigma'_i,\{\sigma_{j\neq i}\})
%w_\mathrm{st}^{(0)}(\sigma'_i,\{\sigma_{j\neq i}\}) w_\mathrm{st}^{(1)}(\sigma'_i,\{\sigma_{j\neq i}\})  =W^{(1)}
\bar W w_\mathrm{st}^{(0)}\sum_{i=1}^N\exp[-\bar K\sigma_i(\sigma_{i+1}+ \sigma_{i-1})]
\left[w_\mathrm{st}^{(1)}(\sigma_i,\{\sigma_{j\neq i}\}) - w_\mathrm{st}^{(1)}(-\sigma_i,\{\sigma_{j\neq i}\})\right] = W^{(1)}
\end{align} 
where %$W_i^{(0)} = \bar W\exp[-\bar K\sigma_i(\sigma_{i+1} + \sigma_{i-1})]$ and 
\begin{align}
\label{eq:W_alternative}
&W^{(1)}% \equiv  - \sum_i\sigma_i \sum_{\sigma'_i}\sigma'_i W_i^{(1)}(\sigma'_i,\{\sigma_{j\neq i}\})
%w_\mathrm{st}^{(0)}(\sigma'_i,\{\sigma_{j\neq i}\})  \nonumber\\
=2 \bar W w_\mathrm{st}^{(0)}\sum_{i=1}^N\exp[-\bar K\sigma_i(\sigma_{i+1}+ \sigma_{i-1})]  \left[\sigma_i (\delta K_{i\,i+1}\sigma_{i+1} + \delta K_{i\,i-1}\sigma_{i-1}) \right]
\nonumber\\
&= -2\bar W w_\mathrm{st}^{(0)}\sinh \bar K\,\sum_{i=1}^N \exp(-\bar K\sigma_i \sigma_{i+1})\delta K_{i\,i+1}\left(\sigma_{i-1}\sigma_{i+1} - \sigma_i\sigma_{i+2} \right)
\end{align}
Here we used that, to zeroth order in the asymmetry, the switching rate is $W_i^{(0)} = \bar W\exp[-\bar K\sigma_i(\sigma_{i+1} + \sigma_{i-1})]$. We note that the factor $w_\mathrm{st}^{(0)}$ drops out from Eq.~(\ref{eq:correction_general}). The set of linear equations (\ref{eq:correction_general}) for $2^{N-1}$ independent components of $w_\mathrm{st}^{(1)}(\{\sigma_i\})$   (on account of the symmetry $\{\sigma_i\} \to \{-\sigma_i\}$) has actually only $2^{N-1}-1$ independent equations and needs to be complemented  by the condition $\sum_{\{\sigma_i\}}w_\mathrm{st}^{(0)}(\{\sigma_i\}) w_\mathrm{st}^{(1)}(\{\sigma_i\}) = 0$.

Equation (\ref{eq:W_alternative}) shows that the correction $W^{(1)}$ can be written as an ordered sum of terms $\propto \delta K_{i\,i+1}$. There are $N$ independent coefficients $\delta K_{i\,i+1} = -\delta K_{i+1\,i}$. 
It immediately follows from Eqs.~(\ref{eq:correction_general}) and (\ref{eq:W_alternative}) that the correction to the stationary distribution, even though linear in $\delta K_{ij}$, is ``nonlocal''. Indeed, one cannot find a spin configuration such that only one term  $\delta K_{i\,i+1}$ is left in $W^{(1)}$. Therefore the effect of the correction is not reduced to modifying just $N$ terms in the distribution.

We note that for $N=2$ we have from Eq.~(\ref{eq:W_alternative}) $W^{(1)}=0$. In this case there is no first-order correction to the stationary distribution. This is seen from the explicit form of $w_\mathrm{st}$ in the main text. However, already for $N=3$, one can see from Eq.~(\ref{eq:correction_general}) that all probabilities $w_\mathrm{st}^{(1)}(\{\sigma_i\})$ are affected by $\delta K_{ij}$.

We note also that the difference of the left- and right-hand sides of Eq.~(\ref{eq:correction_general}) can be written as the sum of the site currents $\sum_i I(\sigma_i,\{\sigma_{j\neq i}\}\to -\sigma_i, \{\sigma_{j\neq i}\}) $. The current on each site is non-zero already in the first order in the asymmetry $\delta K_{ij}$, but their sum of the currents over the sites is equal zero. This is the analog of the Kirchhoff law for the probability current in the stationary regime.

%%%%%%%%%%%%%%%%%%%%%%%%%%%%%%%
%%%%%%%%%%%%%%%%%%%%%%%%%%%%%%%%%

%\section*{Asymmetric Glauber's chain}
%\label{sec:period_two}

%\hfill 

%\setlength{\parindent}{0em}
%\selectfont
\section*{Supplementary Note 10: Asymmetric Glauber chain}
%\setlength{\parindent}{2em}
%\hfill \break

We now consider the dynamics of a chain of weakly coupled spins that describe parametric oscillators. The analysis is simplified in the Glauber limit of weak  nearest-neighbor coupling, $K_{ij}\propto \delta_{|i-j|,1}$ and $|K_{ij}|\ll 1$, in which case 
\[W_i(\sigma_i,\{\sigma_{j\neq i}\}) \approx  \bar W_i \Bigl(1-K_{i\,i+1}\sigma_i \sigma_{i+1}- K_{i\,i-1}\sigma_i \sigma_{i-1}\Bigr).\]
The equation for the mean spin can be obtained by multiplying the kinetic equation (\ref{eq:better_Ising_S}) by $\sigma_i$ and taking trace over all $\{\sigma_j\}$ with the weight given by the probability distribution, as it was done by Glauber.  This gives
\begin{align}
\label{eq:asymmetric_ordered_chain}
\frac{d}{dt}\braket{\sigma_i} = -\bar W_i \Bigl(\braket{\sigma_i}- K_{i\,i+1} \braket{\sigma_{i+1}} - K_{i\,i-1}\braket{\sigma_{i-1}}\Bigr).
\end{align}

The dynamics of the chain is further simplified if the chain is periodic, i.e., every other spin is the same, but the coupling of the spins on even sites to those on odd sites differs from the coupling of the spins on odd sites to those on even sites, 
\[ K_{2i\,2i\pm 1} = K_\mathrm{e}, \, K_{2i+1\, 2i+1\pm 1} = K_\mathrm{o}.\]
We also set the switching rates in the absence of the coupling to be periodic, too,
\[\bar W_{2i} \equiv \bar W_\mathrm{e}, \qquad \bar W_{2i+1} = \bar W _\mathrm{o}, \quad \forall i.\] 

The periodic chain has two spins per unit cell. If the number of cells is $N$ (that is, in Eq.~(\ref{eq:asymmetric_ordered_chain}) $i=1,2,...,2N$ and $\sigma_{i+2N} = \sigma_i$), we can go to the Fourier transform 
\[\sigma_\mathrm{e}(k) = \sum_{n=1}^{N} e^{2ink}\braket{\sigma_{2n}},\quad  
\sigma_\mathrm{o}(k) = \sum_{n=1}^{N} e^{ik(2n-1)}\braket{\sigma_{2n-1}}.\]
The equations for the Fourier components for periodic boundary conditions read
\begin{align}
\label{eq:periodic_chain}
&\frac{1}{\bar W_\mathrm{e}}\frac{d}{dt}\sigma_\mathrm{e}(k) = - \sigma_\mathrm{e}(k) + 2K_\mathrm{e} \sigma_\mathrm{o}(k)\cos k,\nonumber\\
 &\frac{1}{\bar W_\mathrm{o}}\frac{d}{dt}\sigma_\mathrm{o}(k) = - \sigma_\mathrm{o}(k) + 2K_\mathrm{o} \sigma_\mathrm{e}(k)\cos k
 \end{align}

Equations (\ref{eq:periodic_chain}) describe diffusion waves that propagate in the chain. Such waves decay in time, for a given $k$, that is, if we consider a wave $\propto \exp[ikn - i\omega(k) t]$, the ``eigenfrequency'' $\omega(k)$ has an imaginary component. Because there are two spins per unit cell, there are two branches of the diffusion waves. Their eigenfrequencies are 
\begin{align}
\label{eq:diffusion_eigenfrequencies}
\omega_{1,2}(k) =-\frac{1}{2}i(\bar W_\mathrm{e}+ \bar W_\mathrm{o}) \pm \frac{1}{2} i \left[(\bar W_\mathrm{e}-\bar W_\mathrm{o})^2 
 + 16\bar W_\mathrm{e} \bar W_\mathrm{o} K_\mathrm{e}K_\mathrm{o}\cos^2 k\right]^{1/2}.
\end{align}
In the considered limit of small $|K_\mathrm{e,o}|$, we have Im~$\omega_{1,2}(k)<0$, i.e., diffusion waves decay in time in our nonequilibrium system. A qualitative feature of the system is that this decay can be accompanied by temporal oscillations. This occurs if Re~$\omega(k)\neq 0$. In turn, this requires that $K_\mathrm{e}K_\mathrm{o}<0$, i.e.,  $K_{2i, 2in\pm1}$ and $K_{2i+1,2i+1\pm 1}$ have opposite signs. Also, for small $|K_\mathrm{e,o}|$ this requires that the switching rates in the absence of coupling be close, $|\bar W_\mathrm{e} - \bar W_\mathrm{o}|\ll \bar W_\mathrm{e,o}$. As noted earlier, for parametric oscillators the coupling parameters $K_{ij}$ and $K_{ji}$ have the same sign, and therefore the frequencies (\ref{eq:diffusion_eigenfrequencies}) are real.

To the lowest order in the coupling $K_\mathrm{e,o}$ one can find the stationary distribution of the periodic chain and write it as 
\begin{align}
\label{eq:stationary_period_two}
w_\mathrm{st} \approx  Z^{-1}&\exp\left[
\frac{\bar W_\mathrm{e} K_\mathrm{e} + \bar W_\mathrm{o}K_\mathrm{o}}
{\bar W_\mathrm{e} + \bar W_\mathrm{o}}% \right.\nonumber\\
%&\left.\times
\sum_i\sigma_{2i}(\sigma_{2i+1} + \sigma_{2i-1})\right]
\end{align}
To the zeroth order in the coupling the partition function is $Z\to Z_0 = 2^{2N}$. We emphasize that the solution (\ref{eq:stationary_period_two}) only works to the lowest order in $K_\mathrm{e}, K_\mathrm{o}$. Beyond this approximation the distribution $w_\mathrm{st}$ cannot be written in the simple Ising form of the exponential of the sum $\sum_i\sigma_{2i}(\sigma_{2i+1} + \sigma_{2i-1})$.   

Even though the chain is periodic, there is still a nonzero probability current. For example, the current for the flip of an even-site spin while the surrounding spins have the same orientation as the even-site spin in the initial state is 
\begin{align}
\label{eq:current_periodic}
I(\sigma_{2i}=\sigma_{2i\pm 1} = 1 \to \sigma_{2i}=-1, \sigma_{2i\pm 1}=1)% \nonumber\\
=\frac{\bar W_\mathrm{e} \bar W_\mathrm{o}}{2(\bar W_\mathrm{e} + \bar W_\mathrm{o})}(K_\mathrm{o}-  K_\mathrm{e}).
\end{align}
Here  we have taken into account that we have to sum over all spin configurations except for the spins on the sites $2i-1, 2i, 2i+1$. Equation (\ref{eq:current_periodic}) shows that, if spins on the ``odd'' sites are ``stronger'', $K_\mathrm{o} > K_\mathrm{e}$, and are aligned, $\sigma_{2i+1} = \sigma_{2i-1}$, the probability current  flows from the configuration where even-site spins are parallel to odd-site ones to the configuration where they are antiparallel. 

The even-odd configuration can be easily extended to a square lattice. This opens a question of the onset of ``magnetization'' in a large periodic system depending on the asymmetry. Such magnetization would correspond to all parametric oscillators vibrating in phase.

 \end{widetext}
%\end{document}

%%%%%%%%%%%%%%%%%%%%%%%%%%%%%%%%%%
%%%%%%%%%%%%%%%%%%%%%%%%%%%%%%%%%%%%%%%%%%%%

\end{document}